\documentclass[pre,aps,showkeys,showpacs,superscriptaddress,twocolumn,%
amsmath,amssymb]{revtex4} 

\usepackage{graphics}
\usepackage{epsfig}
\usepackage{epstopdf}
\usepackage{amsmath}
\usepackage{setspace}
\usepackage[bookmarks=false,linkcolor=blue,urlcolor=black,colorlinks,citecolor=blue]{hyperref}

\begin{document}
\title{Gaps between avalanches in one-dimensional random field Ising models}

\author{Jishnu N. Nampoothiri} 
\email{jishnu@brandeis.edu}
\affiliation{Martin Fisher School of Physics, Brandeis University, Waltham, Massachusetts 02454, USA}
\author{Kabir Ramola} 
\email{kramola@brandeis.edu}
\affiliation{Martin Fisher School of Physics, Brandeis University, Waltham, MA 02454, USA}
\author{Sanjib Sabhapandit}
\email{sanjib.sabhapandit@gmail.com}
\affiliation{Raman Research Institute, Bangalore - 560080, India}
\author{Bulbul Chakraborty}
\email{bulbul@brandeis.edu}
\affiliation{Martin Fisher School of Physics, Brandeis University, Waltham, MA 02454, USA}

\date{\today} 

\pacs{75.60.Ej, 75.50.Lk, 45.70.Ht}
%75.60.Ej for Barkhausen effect (magnetic properties and materials)
%75.50.Lk for Spin glasses and other random magnets
%45.70.Ht for Avalanches (granular systems)
%61.43.-j for Disordered solids
%63.50.-x for Vibrational states in disordered systems
%63.50.Lm for Glasses and amorphous solids
%64.70.Q- for Theory and modeling of the glass transition
\keywords{Barkhausen Noise, Gap Statistics, Random Field Ising Model}

\begin{abstract}
We analyze the statistics of gaps ($\Delta H$) between successive avalanches in one dimensional random field Ising models (RFIMs) in an external field $H$ at zero temperature. In the first part of the paper we study the nearest-neighbor ferromagnetic RFIM. We map the sequence of avalanches in this system to a non-homogeneous Poisson process with an $H$-dependent rate $\rho(H)$. We use this to analytically compute the distribution of gaps $P(\Delta H)$ between avalanches as the field is increased monotonically from $-\infty$ to $+\infty$. We show that $P(\Delta H)$ tends to a constant $\mathcal{C}(R)$ as $\Delta H \to 0^+$, which displays a non-trivial behavior with the strength of disorder $R$. We verify our predictions with numerical simulations. In the second part of the paper, motivated by avalanche gap distributions in driven disordered amorphous solids, we study a long-range antiferromagnetic RFIM. This model displays a gapped behavior $P(\Delta H) = 0$ up to a system size dependent offset value $\Delta H_{\text{off}}$, and   $P(\Delta H) \sim (\Delta H - \Delta H_{\text{off}})^{\theta}$ as $\Delta H \to H_{\text{off}}^+$. We perform numerical simulations on this model and determine $\theta \approx 0.95(5)$. We also discuss mechanisms which would lead to a non-zero exponent $\theta$ for general spin models with quenched random fields.
\end{abstract}

\maketitle

\section{Introduction}

Many disordered systems when subjected to an external drive, such as a ferromagnet in a magnetic field or a sheared amorphous solid, display a characteristic intermittent response, broadly classified as `crackling noise' \cite{sethna_review_2004,sethna_crackling_review_nature_2001}. This response is characterized by sudden changes in global properties such as magnetization or stress through `avalanches' within the system and can be attributed to the quenched randomness present within these materials. 
The disorder is caused for example, by defects in crystalline solids, by magnetic impurities in the case of spin systems, or the random arrangement of particles in amorphous solids. The properties of avalanches in disordered systems have been of considerable interest in fields ranging from geology to physics \cite{barkhausen_zphys_1919,gutenberg_bul.seis_1944, gutenberg_bul.seis_1956,shcherbakov_prl_2005}.
Various characteristics of avalanches have been investigated including the distribution of their sizes, duration and spatial features \cite{zapperi_prl_2000,sethna_nphys_2011}. Theoretical models such as the well known depinning model successfully describe many key features of crackling noise in these systems \cite{dobrinevski_thesis}. However, developing a general framework with which to describe the response of disordered systems remains an outstanding challenge in the field. Although this response depends non-trivially on the rate of the driving \cite{white_prl_2003}, the limit of infinitesimally slow or `quasi-static' drive is of particular interest.

\begin{figure}
\hspace*{-0.5cm}
\vspace*{-0.5cm}
\includegraphics[width=1.1 \columnwidth]{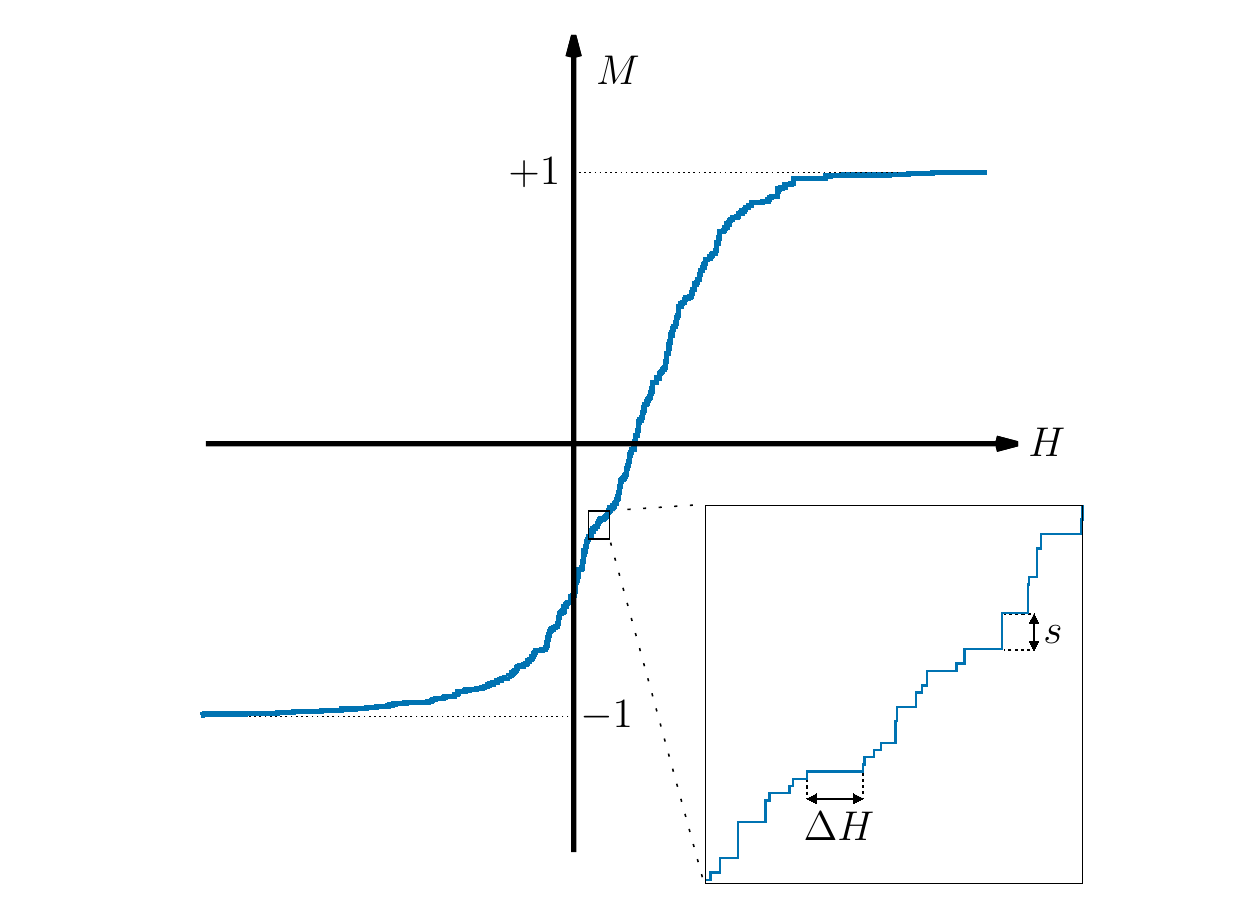}
\caption{The increase in magnetization per site $M$ in the random field Ising model at zero temperature as the external field $H$ is increased monotonically from $-\infty$ to $+\infty$. The jumps in magnetization of size $s$ correspond to avalanches in the system and occur at certain values of the external field $\{H_1 < H_2 < H_3 ...\}$. We study the gaps $\Delta H_{i} = H_{i+1} - H_{i}$ between successive avalanches.}
\label{barkhausen_fig}
\end{figure}

Recent studies of amorphous materials subject to a quasi-static shear have focused attention on another aspect of avalanches in these systems, namely the gaps between successive events \cite{wyart_epl_2014,karmakar_pre_2010}. 
When subjected to increasing strain $\gamma$, amorphous solids undergo stress drops, caused by internal rearrangements. These occur at distinct values $\gamma_1 < \gamma_2 <.. <\gamma_N$ for a given realization of the system. The statistics of these gaps $P(\Delta \gamma)$ with $\Delta \gamma = \gamma_{i+1} - \gamma_i$, yields interesting information about the  stability of the system \cite{wyart_anrev_2015}.
Recently, it has been shown that these `gap statistics' can also be used to distinguish between different phases of such systems \cite{karmakar_pre_2010}.
Crucially, there is a characteristic difference in the statistics of gaps between the process of yielding in amorphous solids and that predicted by the standard depinning process \cite{wyart_epl_2014, wyart_pnas_2014}. This difference is quantified by an exponent $\theta$, defined as $P(\Delta \gamma) \sim \Delta \gamma^{\theta}$ as $\Delta \gamma \to 0$. $\theta$ is always zero in the depinning model but is non-zero in some range of the driving field in amorphous solids. In  jammed packings of frictionless spheres, the exponent $\theta$ can also be related to the distribution of internal forces in the system \cite{wyart_prl_2012}. 

Disordered spin models have been paradigmatic systems to study avalanche behavior \cite{imry_ma_prl_l975,sethna_review_2004}. Many aspects of crackling noise have been well described with models of interacting Ising spins ($S_i = \pm 1$) on a lattice with a quenched random field $\{ h_i \}$ at zero temperature.
 As an external field $H$ is increased quasi-statically from $-\infty$ to $+\infty$, the magnetization per site $M$ changes from $-1$ to $+1$ in discrete steps (see Fig. \ref{barkhausen_fig}).
For a given realization of the random field, these changes in $M$ occur at certain values of the external field $\{H_1 < H_2 < H_3 ...  < H_{N_a}\}$, where $N_a$ represents the total number of avalanches that occur between $-1 < M < 1$ and varies for different realizations. The set $\{H_i\}$ can then be treated as a set of ordered random variables. The distribution $P(\Delta H)$ of the gaps $\Delta H_{i} = H_{i+1} - H_{i}$ is then a statistically interesting quantity that provides information about the internal spin rearrangements. Another related quantity of interest is $P(\Delta H|H)$, the probability that beginning with a configuration at field $H$, $\Delta H$ is the smallest increment required to trigger an avalanche \cite{itamar_sastry_pre_2015}.
Motivated by the avalanche statistics in amorphous solids \cite{wyart_pnas_2014} it is then interesting to ask, under what conditions does a disordered spin model with quenched random fields display a non-zero $\theta$ exponent?

In this paper we study the gap statistics in one dimensional random field Ising models (RFIMs) at zero temperature. 
The outline of the paper is as follows. In Section \ref{distribution_of_gaps_section} we study a RFIM with short-ranged ferromagnetic interactions. We map the sequence of avalanche events in this system to a non-homogeneous Poisson process and use it to derive the distribution of gaps between events. 
In Section \ref{nearest_neighbour_rfim} we study the nearest-neighbor ferromagnetic RFIM which falls into this class of models. Using the above mapping, we compute both the gap distributions $P(\Delta H|H)$ and $P(\Delta H)$ analytically. We show that these distributions tend to constants as $\Delta H \to 0$ for all values of the system parameters, i.e. $\theta = 0$. We verify our predictions with numerical simulations. In Section \ref{long_range_rfim_section}, we study the long-range antiferromagnetic RFIM, that falls outside the class studied in Section \ref{distribution_of_gaps_section}. We perform numerical simulations and use scaling arguments to determine that this model displays a gapped behavior $P(\Delta H) = 0$ up to a system size dependent offset value $\Delta H_{\text{off}}$, and   $P(\Delta H) \sim (\Delta H - \Delta H_{\text{off}})^{\theta}$ as $\Delta H \to H_{\text{off}}^+$. We estimate $\theta \approx 0.95(5)$ independent of model parameters. Finally, in Section \ref{discussion_section} we discuss a possible mechanism which would lead to a non-zero pseudo-gap exponent $\theta$ in this model.

\section{Gaps Between Avalanches in Short-Ranged Ferromagnetic Models}
\label{distribution_of_gaps_section}

In this Section we examine the nature of the distribution of gaps between avalanches in a generic system with short-ranged destabilizing interactions in the presence of quenched disorder. To examine the behavior of avalanches in such systems, we consider a simplified model of $N$ Ising spins $S_i = \pm 1$ in $d$ spatial dimensions. We introduce a ferromagnetic coupling with a finite range $\delta$ between spins, a quenched disorder field $\{h_i\}$ at every site and subject the system to an increasing quasi-static external field $H$. The spins represent the internal state of the constituents of the system, while the ferromagnetic interaction represents a destabilizing interaction between the components, i.e. when an internal restructuring occurs ($-1 \to +1$), it decreases the external field required to restructure the neighboring constituents. The disorder $\{ h_i \}$ is drawn from an underlying distribution $\phi(h,R)$ where $R$ controls the strength of the disorder (typically through the width of the distribution). We  derive a generalized distribution of gaps between avalanche events for such a model using a coarse grained description, essentially treating failures in the system as independent events. This formulation then relates the gap distributions $P(\Delta H|H)$ and $P(\Delta H)$ to the underlying density of failures $\rho_N(H,R)$ in the system. 

\subsection{Mapping to a nonhomogeneous Poisson process}

Consider a realization of the system with a quenched random field $\{h_i\}$, at an external field $H = -\infty$ (i.e. all $S_i = -1$). We are interested in the avalanches that occur in the system as the field is increased monotonically (and quasi-statically) from $H=-\infty$ to $H = +\infty$.
At zero temperature, in the absence of thermal fluctuations, the dynamics is deterministic. We can thus, for a given realization of $\{h_i\}$, group the spins in the system into predetermined clusters that undergo avalanches (failures) together at distinct values of the external field $-\infty < H_1 < H_2 < H_3 ...< +\infty$. 
A key feature of the ferromagnetic interactions is that once a spin flips, it remains in that state. 
Each spin can therefore be uniquely assigned to a cluster. 
This assignment can of course fail for models with stabilizing (such as antiferromagnetic) interactions which we will focus on in Section \ref{long_range_rfim_section}. Every event is initiated at one constituent spin within the cluster and propagates until the entire cluster of spins has flipped. 
Therefore the size of each cluster $s_i$ corresponds to the size of the avalanche event.

\begin{figure}
\centering
\includegraphics[width = 0.9 \columnwidth]{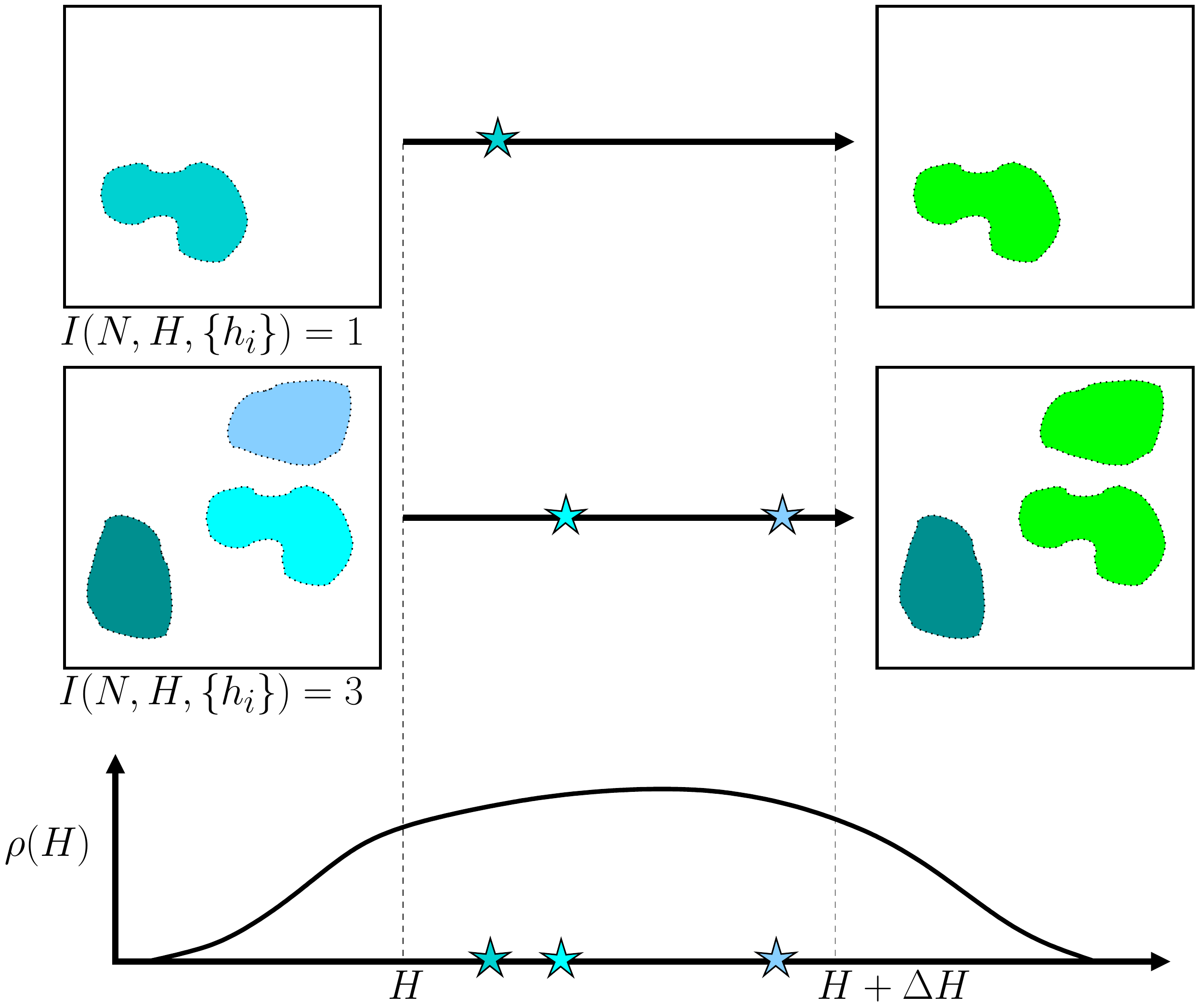}
\caption{A schematic representation of avalanches in the disordered Ising system. On the left are two realizations of the system at a particular value of the external field $H$.  The white region represents spins that have already flipped from $-1$ to $+1$. The coloured areas depict clusters of spins that flip together (avalanche) and have yet to undergo a failure. The number of such regions in each configuration is denoted by $I_N(H,\{h_i\})$, where $N$ is the total number of spins in the system. When the field is incremented by a value $\Delta H$, some of these regions undergo failures at different values of $H$ (represented by stars).  The green regions on the right represent these clusters post-avalanche. In the limit of large $N$, the correlations between events tends to zero, and each of these events can be treated as being independently drawn from an underlying distribution $\rho(H)$.
%The failure of each cluster is assumed to occur independent of each other with a probability $P_{f}(H)$.
}
\label{fig:toy}
\end{figure}

Now when the field is incremented from $-\infty$ to a value $H$, some fraction of the clusters have already undergone failure. We denote the number of clusters yet to undergo a failure at $H$ by $I_N(H,\{h_i\})$, which is a monotonically decreasing function of the field, and serves as a cumulative avalanche density.  This is represented schematically in Fig. \ref{fig:toy}. 

We next argue that for the purposes of analyzing the gap statistics of such models, as long as the interaction range $\delta$ and the average avalanche size $\langle s \rangle$ is finite, the correlations between the avalanche events can be neglected in the thermodynamic limit $N \to \infty$.
In this case, 
the clusters interact only through their boundaries up to a finite distance $\delta$.  Therefore,  events separated by large enough distances in space are uncorrelated with each other. 
Since the events within a given window $[H,H + \Delta H]$ can occur in any part of the system (see Fig. \ref{fig:toy}), it then follows that the probability of events being in close proximity in $H$ {\it and} in space tends to zero as $N \to \infty$ (see Appendix \ref{joint_distribution_section}). Therefore, in the thermodynamic limit, we can essentially treat the avalanches as uncorrelated events.

With this in mind, we consider the ensemble of configurations at different realizations of the quenched disorder at a given $R$ and an external field $H >-\infty$. We define $I_N(H,R)$ to be the average number of clusters which have not failed up to $H$. We then have
\begin{equation}
I_N(H,R) = \langle I_N(H,\{h_i\}) \rangle_{\{h_i \}},
\end{equation}
where the average is taken over all realizations of the quenched disorder.
%We define $f_N(H,R)dH$ as the probability that a cluster fails through an avalanche in any of these configurations when the field is incremented from $H$ to $H+dH$. 
%It is easy to see that
%\begin{equation}
%f_N(H,R) = -\frac{1}{I_N(H,R)} \frac{\partial}{\partial H} I_N(H,R).
%\end{equation}
%This leads us to a coarse grained picture of mutually independent avalanche %events. 
The average density of events at $H$ is then given by 
\begin{equation}
\rho_N(H,R) = -\frac{\partial}{\partial H} I_N(H,R).
\label{rho_definition}
\end{equation}
The mutual independence of failure events now allows us to map the sequence of avalanches in this model to a non-homogeneous Poisson process \cite{book_daley_verejones} with an $H$-dependent rate $\rho_N(H,R)$.
Henceforth for clarity of presentation, we will drop the explicit dependence on $R$ and $N$ of the coarse-grained quantity $\rho_N(H,R)$ (and all subsequent distribution functions derived using it), keeping in mind that $\rho(H) \equiv \rho_N(H,R)$.
We then have
\begin{equation}
\mathcal{N}_1(-\infty,+\infty) = \int_{-\infty}^{\infty}\rho(H) dH = \frac{N}{\langle s \rangle}.
\label{normalization}
\end{equation}
We can then use this to compute the statistics of gaps between avalanches. In Section \ref{nearest_neighbour_rfim} we test the validity of this mapping using simulations of the one-dimensional nearest-neighbor ferromagnetic RFIM.

\subsection{Gap distribution}

The probability of an avalanche occurring at a given value of the external field $H$ is proportional to $\rho(H)$. The probability that successive avalanches occur at field values $H$ and $H'$ can be computed as the joint probability that events occur at $H$ and $H' > H$, with no events between them. This is given by
\begin{equation}
P(H,H') \propto \rho(H)\rho (H') e^{
-\int_{H}^{H'} \rho(y) dy }.
\label{joint_distribiution}
\end{equation}
Testing such a quantity in experiments or simulations, would require conditioning the measurement on an avalanche occurring exactly at $H$, which is a low probability event. Instead, we can focus on a related measure $P(\Delta H|H)$, defined as the probability that starting at a configuration at $H$, the first avalanche occurs at a field increment $\Delta H$. This quantity is sometimes referred to as the instantaneous inter-occurrence time \cite{book_daley_verejones}, and is easier to measure in practice in comparison to $P(H,H')$.
In systems where the gap distribution has different qualitative behaviors at different values of $H$, for example a system which develops long-ranged correlations at some $H_c$, the distribution $P(\Delta H|H_c)$ becomes a more relevant quantity \cite{itamar_sastry_pre_2015}. $P(\Delta H|H)$ can be simply computed as the probability that no avalanche happens in the system when the field is increased from $H$ to $H+\Delta H$ {\it and} an avalanche happens at $H+\Delta H$.
This is given by
\begin{equation}
P(\Delta H|H)= \mathcal{N}_2 \rho (H+\Delta H) \exp\left(
-\int_H^{H+\Delta H} \rho(y) dy \right),
\label{eq:pdh}
\end{equation}
where $\mathcal{N}_2$ is a normalizing factor that ensures $\int_0^{\infty} P(\Delta H|H) d (\Delta H) = 1$ at each $H$, and can be computed to be
\begin{equation}
\mathcal{N}_2^{-1} = 1 - \exp \left( - \int_{H}^{\infty} \rho(y) dy \right).
\end{equation}
For models where the average cluster size $\langle s \rangle$ is finite, it can be seen from Eq. (\ref{normalization}) that the integral in the exponential in Eq. (\ref{eq:pdh}) scales as $N$, the total number of spins. In Section \ref{nearest_neighbour_rfim}, we measure this distribution in detail for the one dimensional nearest-neighbor ferromagnetic RFIM using numerical simulations, and compare it to an analytic expression derived using Eq. (\ref{eq:pdh}).

We can next use the expression in Eq. (\ref{eq:pdh}) to investigate the pseudo-gap exponent $\theta$ for $P(\Delta H| H)$. The expression in  Eq. (\ref{eq:pdh}) in the small $\Delta H$ regime can be simplified to
\begin{align}
P(\Delta H|H)\sim  \mathcal{N}_2 \rho (H+\Delta H) \exp \left(-\rho(H) \Delta H \right).
\label{small_deltaH_behaviour}
\end{align}
From this we see that the small $\Delta H$ behavior of $P(\Delta H| H)$ is completely governed by the behavior of the $\rho(H)$. If the density of avalanches at some $H_c$ is zero {\it and} has a behavior $\rho(H_c+\Delta H) \sim \left(\Delta H\right)^\theta$ as $\Delta H \to 0^{+}$ in its vicinity,  $P(\Delta H| H)$ would also exhibit a non-zero $\theta$ exponent. 
It is therefore worthwhile to study models where one can compute the density of avalanches exactly.
In Section \ref{nearest_neighbour_rfim} we study the one-dimensional nearest-neighbor ferromagnetic RFIM, where we use the techniques developed in \cite{sanjib_thesis} along with the formalism developed in this section to compute $P(\Delta H| H)$ exactly.

Finally, we consider the distribution of gaps between avalanches in the entire sweep of the magnetic field from $H = -\infty$ to $+\infty$, which is a quantity that is accessible in typical experimental observations. This is given by the expression (see Appendix \ref{gap_distribution_appendix})
\begin{equation}
P(\Delta H) = \int_{-\infty}^{+\infty} \frac{ \rho(H')\rho (H'+\Delta H)}{\mathcal{N}_1(-\infty,+\infty)} e^{-\int_{H'}^{H'+\Delta H} \rho(y) dy } d H',
\label{full_sweep_pdf}
\end{equation}
$\mathcal{N}_1(-\infty,+\infty) = N/\langle s \rangle$ is the normalization defined in  Eq. (\ref{normalization}).
It is then straightforward to extract the small $\Delta H$ behavior from this expression. We have
\begin{equation}
\mathcal{C}(R) = \lim_{\Delta H \to 0}P(\Delta H) = \frac{\int_{-\infty}^{+\infty} \rho(H')^2 d H'}{\int_{-\infty}^{+\infty} \rho(H') d H'}.
\label{Rconstant}
\end{equation}
As long as $\rho(H)$ is finite in a finite range of $H$, $P(\Delta H)$ saturates to a constant $\mathcal{C}(R)$ as $\Delta H \to 0$.
Therefore, we conclude that the pseudo-gap exponent $\theta=0$ for $P(\Delta H)$ in this class of systems. In Section  \ref{nearest_neighbour_rfim} we analyze the one-dimensional nearest neighbor ferromagnetic RFIM and show that the predictions for the gap distributions $P(\Delta H|H)$ and $P(\Delta H)$ from our theory agree well with the results from simulations. As this model falls into the class considered in this section, we verify that the pseudo-gap exponent $\theta=0$ in this case.
Finally, it is clear from the form of Eq. (\ref{full_sweep_pdf}) and using the fact that $\rho(H) \sim N$ from Eq. \eqref{normalization} that as $N \to \infty$, the gap distribution $P(\Delta H )$ has the scaling form
\begin{equation}
P(\Delta H) =  N \mathcal{P} (N \Delta H).
\label{scaling_form}
\end{equation}
Our treatment of avalanches as mutually independent events leads to the conclusion that in order for a system to display a non-zero $\theta$ exponent either in $P(\Delta H|H)$ or $P(\Delta H)$, some of the assumptions made in the above model must fail. 
%For $P(\Delta H|H)$, the assumption that fails is that $\rho(H+\Delta H)$ does not go as $(\Delta H)^{\theta}$, which is an additional assumption beyond the Poisson process ?
This can occur in any number of ways, the correlations between clusters can become long-ranged, the interactions themselves can have a long-ranged component, or there can be stabilizing interactions in the system. In Section \ref{long_range_rfim_section} we construct a long-ranged antiferromagnetic model that has two of these features, and we find that indeed, beyond a system size dependent offset value $\Delta H_{\text{off}}$, this system displays $P(\Delta H) \sim (\Delta H - \Delta H_{\text{off}})^{\theta}$, with $\theta = 0.95(5)$.

\section{The Nearest-Neighbor Ferromagnetic RFIM}
\label{nearest_neighbour_rfim}

In this Section we analyze the properties of the nearest-neighbor ferromagnetic Random Field Ising Model at zero temperature. This model has been successfully used to describe the  noisy response of ferromagnets to external fields \cite{sethna_prl_1993, sethna_review_2004}, which was observed experimentally by Barkhausen \cite{barkhausen_zphys_1919}. In contrast to that of the nearest neighbor ferromagnetic Ising Model where long-range order occurs for $d > 1$, the presence of arbitrarily small disorder destroys long-range order in $d \leq 2$ \cite{grinstein_prb_1983}. 

%Previous studies have investigated the statistics of the avalanche sizes $P(s)$ during the hysteritic response [...]. In this study we focus on the gap statistics of avalanches $P(\Delta H)$. 

The ferromagnetic RFIM has several intriguing properties, such as a no-crossing property \cite{middleton_prl_1992}, an Abelian property and a return point memory \cite{dhar_jphysa_1997}, that make it theoretically accessible \cite{sanjib_jstatphys_2000,sanjib_thesis}. 
For the nearest neighbor RFIM on a Bethe lattice it is indeed possible to compute the probability of an avalanche of size $s$ originating from a given site $P(s,H)$ exactly \cite{sanjib_thesis}. 
It is easy to see that one can then compute the coarse grained density of avalanche events $\rho(H)$.
Defining the generating function  $G(x,H)= \sum_{s = 1}^{\infty}P(s,H)  x^s$ (see Appendix \ref{generating_function_appendix}),
the probability of an avalanche of any size originating from a given spin is simply $G(x = 1,H)$. We therefore obtain
\begin{equation}
\rho(H) = N G(1,H).
\label{eq:rho_to_G}
\end{equation}
Then, using the formalism developed in Section \ref{distribution_of_gaps_section} we can derive the distribution of gaps between avalanches from Eqs. (\ref{eq:pdh}), (\ref{full_sweep_pdf}) and (\ref{eq:rho_to_G}). We compute these distributions for two cases with quenched random fields chosen from 
(i) a bounded distribution (which we choose as a uniform distribution) and (ii) an unbounded distribution (which we choose as an exponential). We show that these two cases have qualitatively different behaviors for the gap distribution. We also numerically simulate this model and show a very good agreement between our theoretical predictions and those obtained from simulations.
 
 The Hamiltonian of the system is given by 
\begin{equation}
\mathcal{H}=-J \sum_{\langle i,j \rangle} S_i S_j - \sum_i \left(h_i+H\right) S_i,
\end{equation}
where $J>0$ represents the ferromagnetic coupling between nearest neighbor spins on the one-dimensional chain, $H$ represents the external magnetic field. $\{h_i\}$ represents the quenched random field at every site, chosen from a distribution $\phi(h,R)$, where $R$ controls the strength of the disorder.
The system evolves under the zero-temperature Glauber single-spin-flip dynamics, i.e. a spin flip occurs only if it lowers the energy.
This is achieved by making each spin align with its effective local field $h_{e,i}$ given by
\begin{equation}
h_{e,i}=J(S_{i-1} + S_{i+1}) + h_i+H .
\label{effective_field_RFIM}
\end{equation}
The system is then relaxed until a stable configuration is obtained at that value of the field $H$, which in the zero temperature dynamics is simply determined by the condition
\begin{equation}
S_i=\text{sign}(h_{e,i}).
\label{energy_minizing}
\end{equation}

 \begin{figure}
  \centering
  \includegraphics[width = 1 \columnwidth]{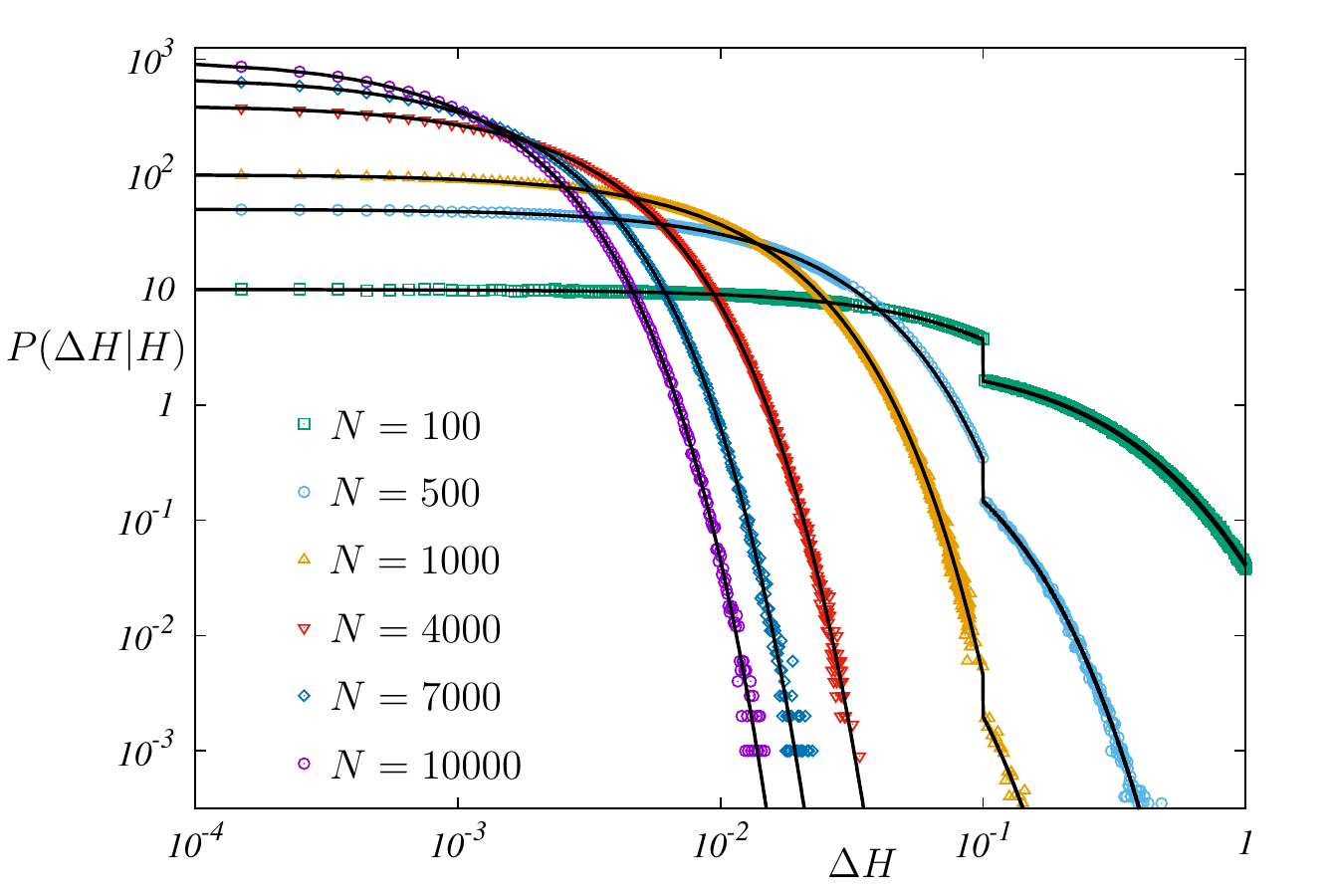}
  \caption{Distribution of gaps between avalanches $P(\Delta H|H)$ in the nearest neighbor ferromagnetic RFIM with uniform disorder at $R=5$ and $H=2.9$ for $10^7$ realizations of the disorder. The bold lines represent analytical results computed using Eqs. (\ref{eq:pdh}), (\ref{eq:rho_to_G}) and (\ref{nn_uniform_g1}). The points represent data obtained from simulations. We find a very good agreement between our analytical results and those obtained from the simulations. The discontinuity in the distribution at occurs at $\Delta H = 2 J - R - H = 0.1$ (Eq. (\ref{nn_uniform_g1})), and reflects the discontinuity in the underlying disorder distribution.}
    \label{uniform_disorder_RFIM_fig2}
\end{figure}
 
%In the appendix \ref{appendix} we derive the generating function for the distribution of avalanche sizes $G(x,H)$ for a general disorder, r. 
%We find that in the case of the linear chain ($z=2$ Bethe lattice), the equations determining $M(H)$ are linear, hence there is no discontinuity in the magnetization as the field is increased. 
We use this dynamics to analyze the generic features of the gap distributions $P(\Delta H|H)$ and $P(\Delta H)$ for two cases of the distribution of quenched random fields $\phi(h,R)$ (i) a uniform distribution and (ii) an exponential distribution
In the first case, $\{h_i\}$ is chosen from a uniform distribution with a width $R$ as
\begin{equation}
 \phi(h,R) =
  \begin{cases} 
      \hfill \frac{1}{2R}    \hfill & |h| \leq R,\\
      \hfill 0 \hfill & |h|>R,\\
  \end{cases}
  \label{uniform_disorder}
\end{equation}
and in the second case, $\{h_i\}$ is chosen from an exponential distribution with a width $\sqrt{2} R$ as
\begin{equation}
\phi(h,R)=\frac{1}{2R}\exp\left(-\frac{|h|}{R}\right).
\label{exponential_disorder}
\end{equation}
In both cases, $R$ is the parameter that controls the strength of the disorder by controlling the width of the distribution $\phi(h,R)$.

\begin{figure}
\centering
\includegraphics[width = 1 \columnwidth]{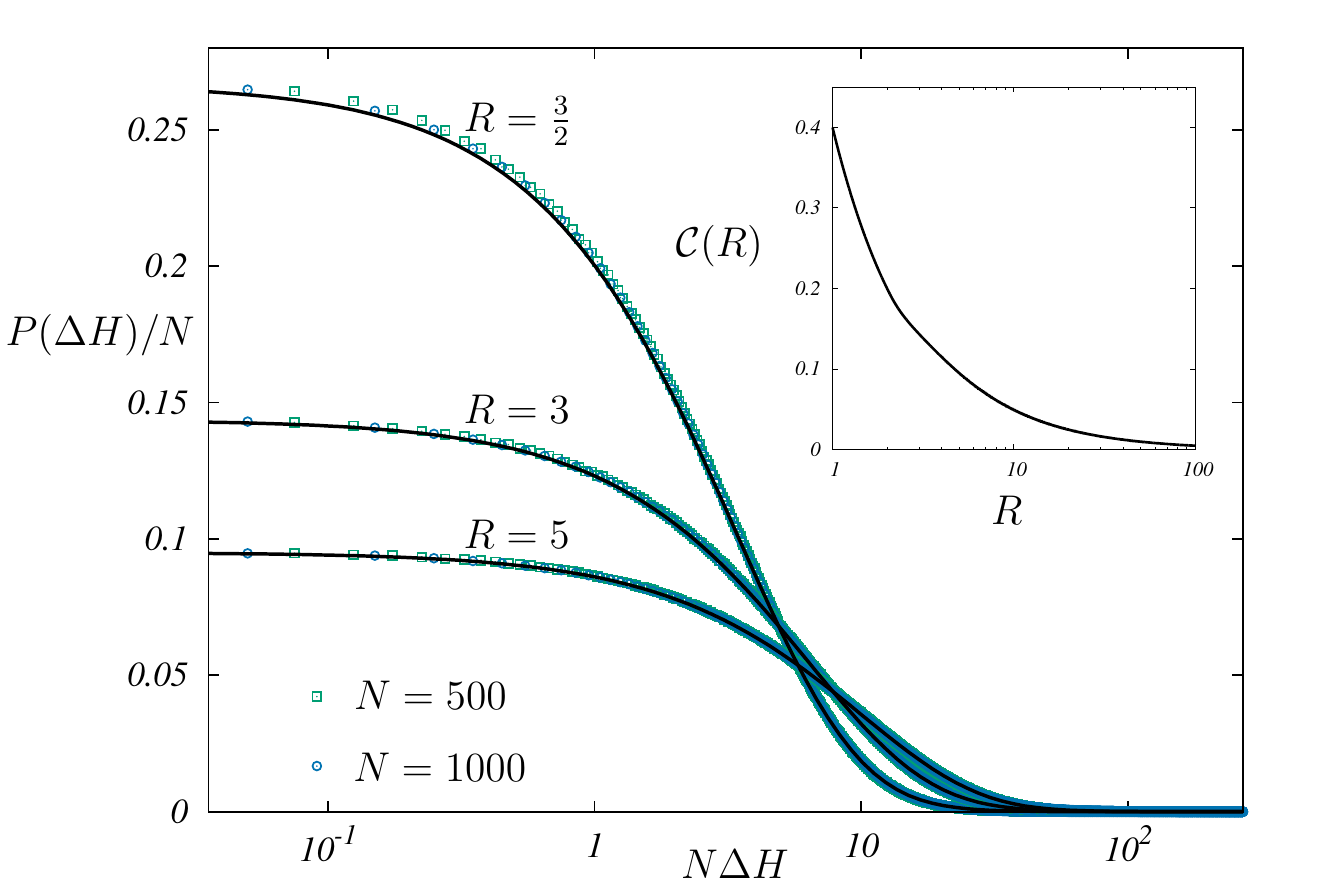}
\caption{Distribution of gaps between avalanches $P(\Delta H)$ in the nearest neighbor ferromagnetic RFIM with uniform disorder for different R. The bold lines represent analytical results computed using Eqs. (\ref{full_sweep_pdf}), (\ref{eq:rho_to_G}) and (\ref{nn_uniform_g1}). The points represent data obtained from simulations. The data have been averaged over $10^7$ realizations. We find a very good agreement between our analytical results and those obtained from the simulations. (Inset) The saturation value $\mathcal{C}(R)  = \lim_{\Delta H \to 0^+}P(\Delta H)$ for different values of $R$.}
  \label{uniform_disorder_RFIM_fig3}
\end{figure}

\subsection{Uniform disorder}

For the case of uniform disorder, given by Eq. (\ref{uniform_disorder}), we find three different regimes depending on the relative strengths of the disorder $R$ and the interaction $J$. When $R<J$, there is a single system sized avalanche which occurs at $H=2J-R$, and the magnetization per site jumps from $M = -1$ to $M = +1$. The other two cases are when $R>2J$ and $J<R\leq 2J$. Here there are several avalanches with a distribution of sizes at different field strengths $H$. 
There is, however, a qualitative difference in the nature of avalanches for the cases $R>2J$ and $J<R\leq 2J$ \cite{sanjib_thesis}.
The form of $G(1,H)$ for these two cases is given by (see Appendix \ref{generating_function_appendix})
\begin{equation}
 G(1,H) =
  \begin{cases} 
      \hfill 0 \quad   \hfill & {\scriptstyle H \leq 2J-R}, \\ 
      \hfill \frac{1}{2R} \quad \hfill & {\scriptstyle 2J-R < H < -2J+R\  \&\  R>2J} ,\\ \\
      \hfill \frac{(R-H)(H+3R-4J)}{8R(R-J)^2} \hfill & {\scriptstyle 2J-R<H<R\ \&\ J<R\leq 2J} ,\\ \\
       \hfill \frac{(R-H)(H+3R-4J)}{8R(R-J)^2} \hfill & {\scriptstyle -2J+R<H<R\  \&\  R>2J} ,\\ \\
       \hfill 0 \quad   \hfill & {\scriptstyle H > R} .\\

  \end{cases}
  \label{nn_uniform_g1}
\end{equation}

\begin{figure}
  \centering
  \includegraphics[width = 1 \columnwidth]{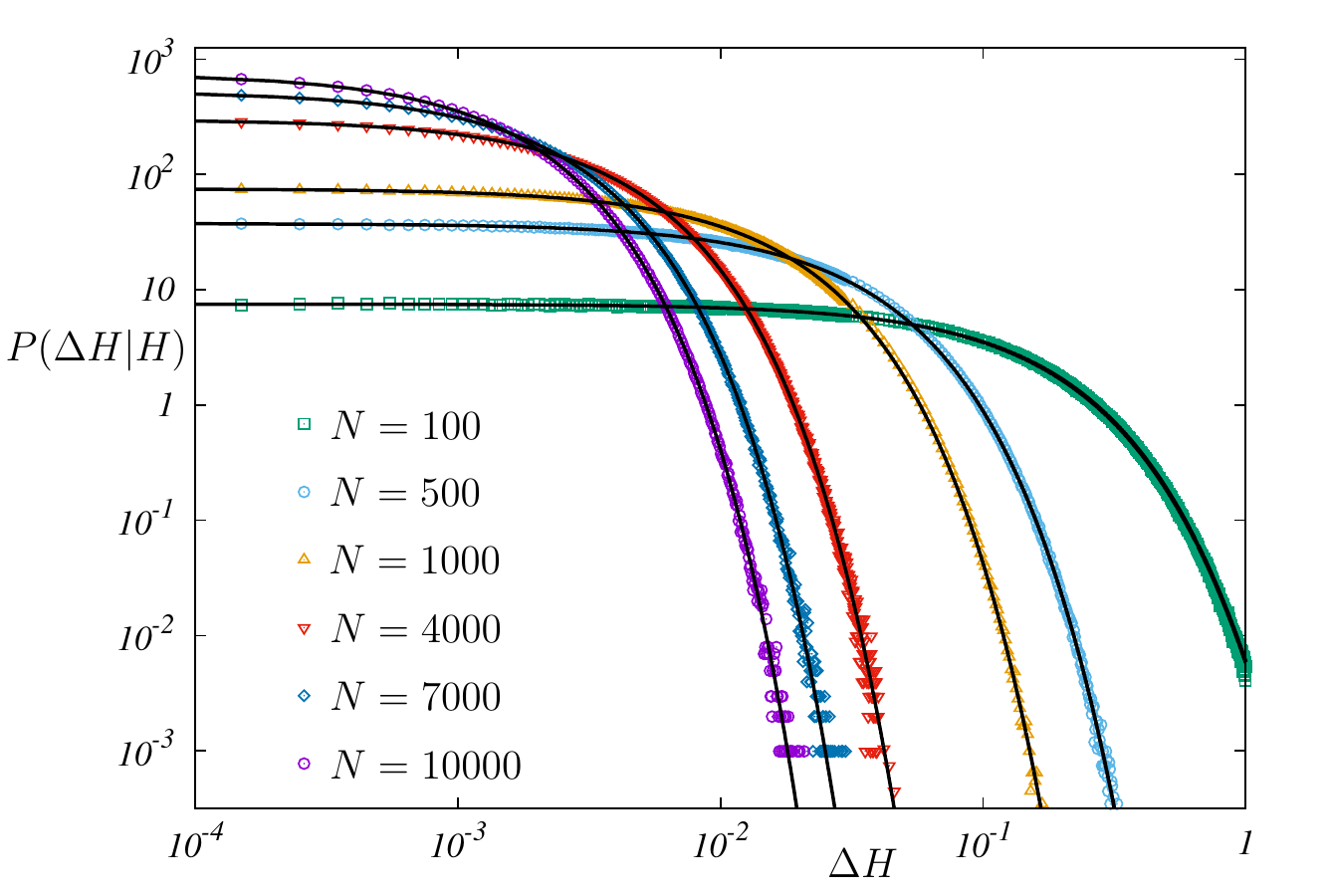}
  \caption{Distribution of gaps between avalanches $P(\Delta H|H=1)$ in the nearest neighbor ferromagnetic RFIM with exponential disorder at $R=5$ and $H=1$ for $10^7$ realizations of the disorder. The bold lines represent analytical results computed using  Eqs. (\ref{eq:pdh}), (\ref{eq:rho_to_G}) and (\ref{nn_exponential_g1}). The points represent data obtained from simulations. The data have been averaged over $10^7$ realizations. We find a very good agreement between our analytical results and those obtained from the simulations.}
  \label{exponential_disorder_RFIM_fig2}
\end{figure}

Eq. (\ref{nn_uniform_g1}) can be used directly to calculate $\rho(H)$ ( Eq. (\ref{eq:rho_to_G})), which in turn allows us to compute the distribution of gaps between avalanches $P(\Delta H| H)$ using Eq. (\ref{eq:pdh}).
As argued in Section \ref{distribution_of_gaps_section}, the small $\Delta H$ behavior of $P(\Delta H|H)$ is controlled by the small $\Delta H$ behavior of the density $\rho(H)$ and consequently $G(1,H)$. Expanding Eq. (\ref{nn_uniform_g1}) we find
\begin{equation}
P(\Delta H|H) \sim \Delta H^{0} ~~~\text{as}~~~ \Delta H \to 0^+, ~~~\forall~~~ H
\end{equation}
leading to a zero pseudo-gap exponent for $P(\Delta H|H)$  in this case. 
This result can be understood as follows, from the arguments in Section \ref{distribution_of_gaps_section}, the presence of a non-zero $\theta$ exponent for $P(\Delta H|H)$ requires that $G(1,H)$ vanish at some $H_c$ {\it and} have a behavior of the form $G(1,H_c+\Delta H) \sim (\Delta H)^\theta$ as $\Delta H  = (H - H_c) \to 0^+$. The only points at which $G(1,H)$ vanishes are $H = R$ and $H = 2 J - R$. $G(1,H)$ is identically zero for all $H > R$, and jumps discontinuously from $0$ to a finite value at $H = 2 J - R$, leading to $\theta = 0$ for $P(\Delta H|H)$ for all $H$, for the case with uniform disorder.
In Fig. \ref{uniform_disorder_RFIM_fig2} we plot  $P(\Delta H|H)$ for $H/J = 2.9$ computed using Eqs. (\ref{eq:pdh}), (\ref{eq:rho_to_G}) and (\ref{nn_uniform_g1}) for the uniform disorder distribution with $R =5$. The discontinuities which appear in the gap distribution in Fig. \ref{uniform_disorder_RFIM_fig2} are purely due to the fact that the underlying disorder distribution $\phi(h,R)$ has a discontinuity. In the next section, we indeed observe that these discontinuities are absent for the case with a continuous (exponential) distributed disorder.

\begin{figure}[t!]
  \centering
  \includegraphics[width = 1 \columnwidth]{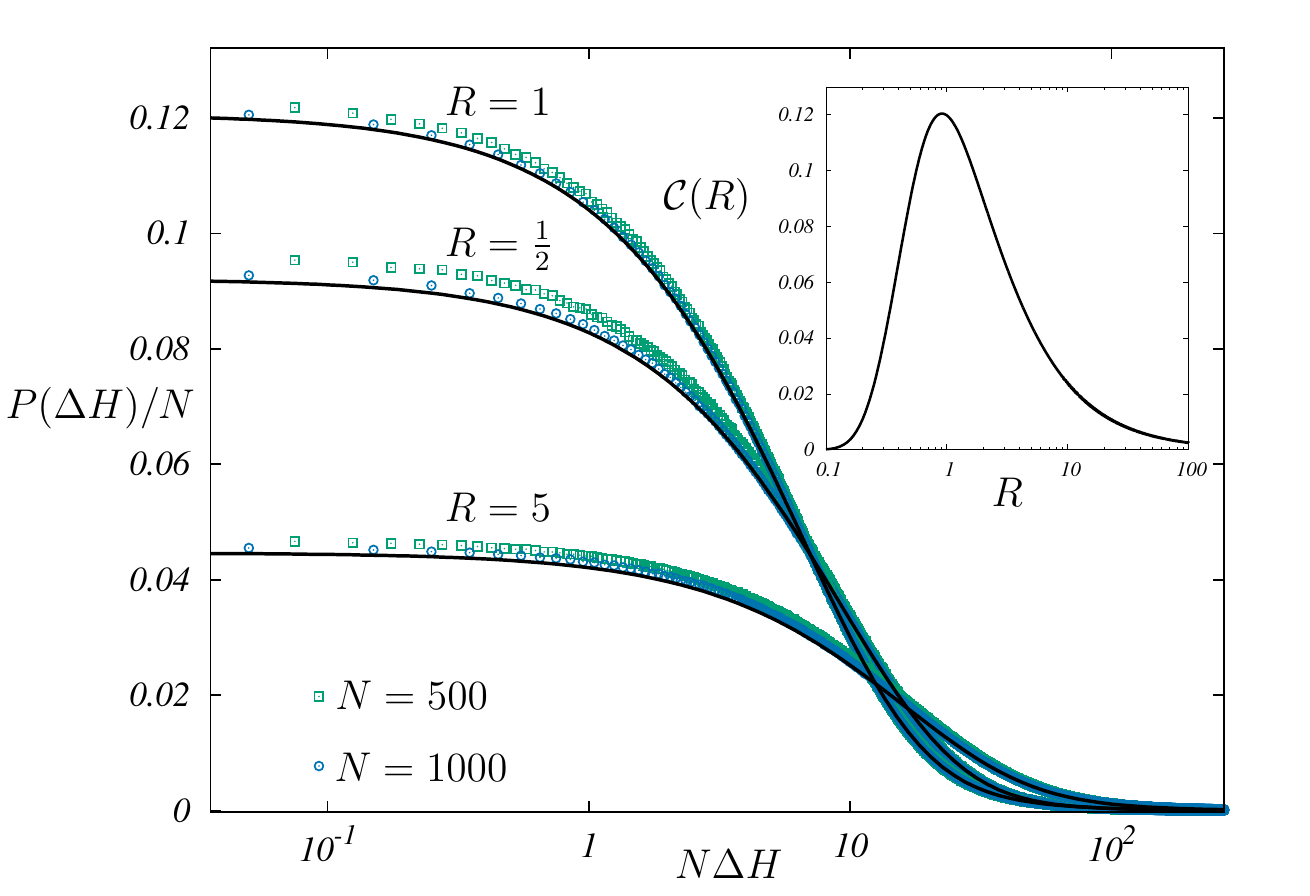}
  \caption{Distribution of gaps between avalanches $P(\Delta H)$ in the nearest neighbor ferromagnetic RFIM with exponentially distributed random fields for different $R$. The bold lines represent analytical results computed using Eqs. (\ref{eq:rho_to_G}), (\ref{full_sweep_pdf}) and (\ref{nn_exponential_g1}). The points represent data obtained from simulations. The data have been averaged over $10^7$ realizations. We find a very good agreement between our analytical results and those obtained from the simulations. (Inset) The saturation value $\mathcal{C}(R)  = \lim_{\Delta H \to 0^+}P(\Delta H)$ for different values of $R$.}
  \label{exponential_disorder_RFIM_fig3}
\end{figure}

Next, we compute the distribution of gaps $P(\Delta H)$ evaluated over an entire sweep in the magnetic field.
In Fig. \ref{uniform_disorder_RFIM_fig3} we plot $P(\Delta H)$ computed using Eqs. (\ref{full_sweep_pdf}), (\ref{eq:rho_to_G}) and (\ref{nn_uniform_g1}) for various values of the disorder strength $R$ at different system sizes. We find that this obeys the scaling form provided in Eq. (\ref{scaling_form}). This distribution saturates to a constant $\mathcal{C}(R)$ as $\Delta H \to 0$, which can be computed using Eqs. (\ref{Rconstant}) and (\ref{eq:rho_to_G}). We show the behavior of $\mathcal{C}(R)$ as a function of $R$ in the inset of Fig. \ref{uniform_disorder_RFIM_fig3}. We find that $\mathcal{C}(R)$ reaches a value $2/5$ as $R \to 1^+$, and decays as $R$ increases. The region below $R < J$, is inaccessible as the system displays a single system sized avalanche at $H = 2J-R$.

\subsection{Exponential disorder}
For the case of an exponentially distributed disorder given by Eq. (\ref{exponential_disorder}), the form of $G(1,H)$ is given by (see Appendix \ref{generating_function_appendix}) 

\begin{equation}
 G(1,H) =
  \begin{cases} 
      \hfill \frac{4 e^{\frac{2J+H}{R}}-e^{\frac{2J+3H}{R}}+e^{\frac{2H-|H+2J|}{R}}}{2R\ \left(2 e^{\frac{2J}{R}}+e^{\frac{H}{R}}-e^{\frac{2J+H}{R}}\right)^2} \quad   \hfill & H < 0, \\ \\
\hfill \frac{3 e^{\frac{2J+H}{R}}+e^{\frac{3H-2J}{R}}}{2R\ \left(e^{\frac{2J}{R}}+e^{\frac{2H}{R}}\right)^2} \quad   \hfill & 0 \leq H \leq 2J, \\ \\

\hfill \frac{2 \ e^{\frac{H-2J}{R}}}{R \  \left(2e^{\frac{H}{R}}+1-e^{\frac{2J}{R}}\right)^2} \quad   \hfill & H > 2J.\\

  \end{cases}
  \label{nn_exponential_g1}
  \end{equation}  
In contrast to the bounded uniform distribution, there are no discontinuities in the gap distribution, since there are no discontinuities in the distribution $\phi(h,R)$. In Fig. \ref{exponential_disorder_RFIM_fig2} we plot $P(\Delta H|H)$ computed using  Eqs. (\ref{eq:pdh}), (\ref{eq:rho_to_G}) and (\ref{nn_exponential_g1}) for the exponential disorder distribution with $R =5$ at a field strength $H/J = 1$ for various system sizes. In this case, $G(1,H)$ is finite everywhere, and therefore from the arguments of Section \ref{distribution_of_gaps_section}, once again $\theta  = 0$ for $P(\Delta H|H)$ for all $H$ in this case.
In Fig. \ref{exponential_disorder_RFIM_fig3} we plot $P(\Delta H)$ computed using Eqs. (\ref{full_sweep_pdf}), (\ref{eq:rho_to_G}) and (\ref{nn_exponential_g1}) for various values of the disorder strength $R$ at different system sizes. We find that this distribution obeys the scaling form given in Eq. (\ref{scaling_form}).
% In this case as well, we find that this obeys the scaling form provided in Eq. (\ref{scaling_form}). 
This distribution once again saturates to a constant $\mathcal{C}(R)$ as $\Delta H \to 0$, which can be computed using Eqs. (\ref{eq:rho_to_G}) and (\ref{Rconstant}). We show the behavior of $\mathcal{C}(R)$ as a function of $R$ in the inset of Fig. \ref{exponential_disorder_RFIM_fig3}. 
Unlike the uniform distribution,  we are able to access the very low disorder regions  $R \ll \sqrt{2} J$, and probe its properties. 
We find that $\mathcal{C}(R)$ displays an intriguing non-monotonic behavior around the point $R \sim  J$. In the high disorder regime, $\mathcal{C}(R)$ decays to $0$ exponentially as $R \to \infty$. In the low disorder regime, it decays to $0$ with an essential singularity as $R \to 0^+$.

\subsection{Numerical simulations}

To test the predictions made by our theory, we perform numerical simulations.
We generate a particular realization of the quenched random field ($\{h_i\}$), drawn from the disorder distribution $\phi(h,R)$. We start from a configuration in which all the spins in the lattice are $-1$, corresponding to $H=-\infty$. The spins are then relaxed to their stable configuration at a given $H$ using single spin flip energy minimizing dynamics (Eqs. (\ref{effective_field_RFIM}) and (\ref{energy_minizing})). Once the spins are relaxed, the smallest increment in the external field required to flip a spin from this stable configuration is computed ($\Delta H$). The field is then incremented to this value ($H+\Delta H$) and the spins are once again relaxed to their stable configuration. The statistics of these increments are used to compute $P(\Delta H|H)$. The avalanche size $s$ is defined as the number of spins which change their state as the field is increased from $H$ to $H+\Delta H$. We repeat this procedure for several realizations of the disorder to generate a distribution of avalanche sizes and gaps at a given $H$. Finally we compute $P(\Delta H)$, by performing a full sweep in $H$ from $-\infty$ to $+\infty$. Our simulations are carried out with periodic boundary conditions and the units are chosen so that $J=1$.
%The simulations are carried out using  MATLAB\textsuperscript{\textregistered}.
We compare the distributions obtained from the theory and simulations in Figs. \ref{uniform_disorder_RFIM_fig2}, \ref{uniform_disorder_RFIM_fig3}, \ref{exponential_disorder_RFIM_fig2} and \ref{exponential_disorder_RFIM_fig3}.

\vspace{1cm}

In summary, we find a very good agreement between our theory and simulations in all regions of the parameter space for both $P(\Delta H|H)$ and $P(\Delta H)$, verifying our analysis of Section \ref{distribution_of_gaps_section}. Our exact results show that the pseudo-gap exponent $\theta$ is zero for the RFIM in one dimension. Although this follows naturally from the fact that the RFIM can be mapped onto a depinning process \cite{grinstein_prb_1983,bruinsma_prl_1984} which is known to have a zero pseudo-gap exponent \cite{wyart_epl_2014}, we have been able to analytically compute this.
The question we next seek to address is the following : What kind of physical interactions in a random field model can give rise to a non-zero $\theta$ exponent for $P(\Delta H | H)$ or $P(\Delta H)$. For $P(\Delta H | H)$ to display a non-zero $\theta$ at some value of the field $H_c$, we require a vanishing of the avalanche density $\rho(H_c)$. Thinking physically, this can happen if the avalanche that occurred prior to the system reaching $H_c$ renders all other regions further from failure, i.e. this avalanche affects a thermodynamically large region of the system. 
%From the discussion in Section \ref{distribution_of_gaps_section}, for $P(\Delta H)$ to display a non-zero $\theta$ exponent, some of the assumptions of our analysis must fail. 
% this can occur at a critical point, or in a system with long-range interactions.  
 From an often used random walk picture of yielding \cite{wyart_pnas_2014},   the interaction needs to have a stabilizing component for an avalanche to render regions further from failure.  A thermodynamically large region can be affected in  a system with long-range interactions or at a critical point for a system with short range interactions.
% and from our analysis of avalanches in short ranged ferromagnetic spin systems, we conclude that the interaction should have a stabilizing component to display a non-zero pseudo-gap exponent. 
Amongst spin systems with random interactions, it is known that spin glasses have non-zero $\theta$ exponents \cite{wyart_anrev_2015}. Since we are focused on random-field models, a disordered spin model with long-range antiferromagnetic interactions is a good candidate for a spin model with a non-zero $\theta$ exponent \cite{travesset_prb_2002}.

\section{The Long-Range Antiferromagnetic RFIM}
\label{long_range_rfim_section}

In this Section we study a long-range antiferromagnetic RFIM that displays a gapped behavior $P(\Delta H) = 0$ up to a system size dependent offset value $\Delta H_{\text{off}}$, and   $P(\Delta H) \sim (\Delta H - \Delta H_{\text{off}})^{\theta}$ as $\Delta H \to H_{\text{off}}^+$.
We consider $N$ Ising spins on a one dimensional lattice. The Hamiltonian of the system is given by
\begin{equation}
\mathcal{H}= J_0 \sum_{i = 1}^{N} \sum_{\substack{j = 1\\j \ne i}}^{N} \frac{S_i S_j}{|i-j|^{1+\alpha}} -\sum_{i=1}^{N}(h_i+H)S_i .
\label{long_range_hamiltonian}
\end{equation}  
Here $J_0 > 0$ represents the antiferromagnetic interaction between the spins. Once again $H$ represents the external field and $\{h_i\}$ represents the quenched random field at every site chosen from a distribution $\phi(h,R)$. We consider exponentially distributed random fields governed by the distribution given in Eq. (\ref{exponential_disorder}). $\alpha>0$ controls the range of interaction in the system. The limit $\alpha \to \infty$ yields the short-ranged antiferromagnetic RFIM. In the limit $\alpha \to 0$ and fixed magnetization per site $M$, this Hamiltonian can be exactly mapped onto the Hamiltonian of the Coulomb glass \cite{andresen2016charge,mapping}.

%%%%%%%%%%%%%%%%%%%%%%%%%%%%%%%%%%%%%%%%%%%%%%%%%%%%%%%%%%%%%
\begin{figure}
\includegraphics[width = 1 \columnwidth]{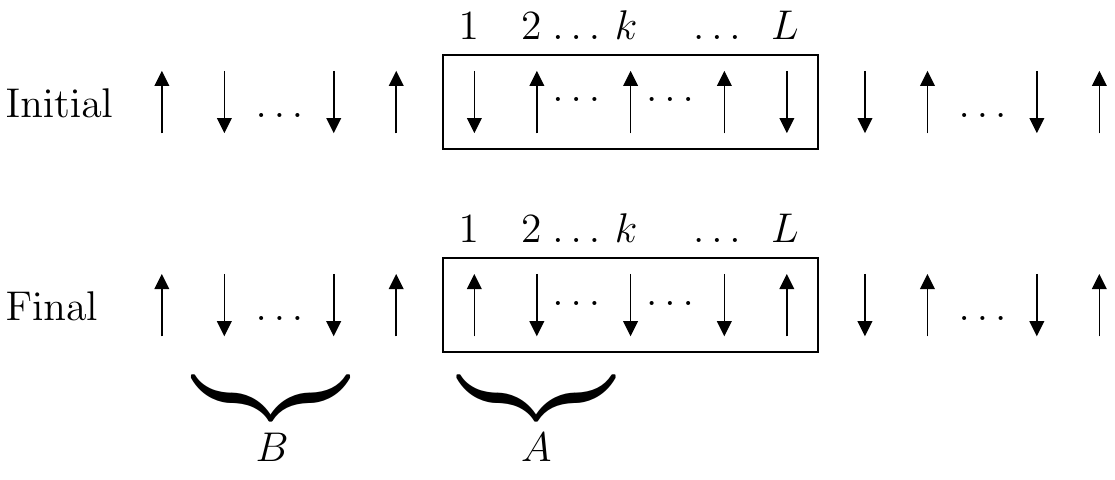}
\caption{A schematic representation of states in the spin model (up(down) arrows correspond to $S_i = +1(-1)$). Starting from the ground (N\'eel) state  (Initial), a block of spins numbered $1,2 \hdots L$ is flipped.
The $A$ term represents the interaction of a spin $k$ with the spins to its left within the block and $B$ represents the interaction of this spin with spins to its left outside the block. The cost of flipping this block of spins can be made arbitrarily small as $L \to \infty$, for any finite disorder. 
\label{fig:imry-ma_rfafim}}
\end{figure}
%%%%%%%%%%%%%%%%%%%%%%%%%%%%%%%%%%%%%%%%%%%%%%%%%%%%%%%%%%%%%

This system has no frustration and has two well defined ground states in the zero disorder, zero external field limit, namely the staggered antiferromagnetic ground (N\'eel) states. This long-range order is destroyed in the presence of any disorder \cite{kerimov_jstatphys_1993}, which we show using an Imry-Ma type argument in Section \ref{absence_of_order_section}. 
Due to the antiferromagnetic nature of the interaction, every spin prefers to be anti-aligned with every other spin in the system. Therefore when the driving field causes a spin to flip from $-1$ to $+1$, this stabilizes all the other spins in the lattice, rendering them further from failure. 

Since the interaction is antiferromagnetic, it is possible for spins to flip back (i.e. a spin goes from being aligned to the external field to anti-aligned), in contrast to the ferromagnetic case. Therefore, it is not possible to uniquely group the spins into clusters that undergo avalanches together as was done in the analysis in Section \ref{distribution_of_gaps_section}. When this system is subjected to an external field, there can be spin rearrangements which do not change the magnetization. It is therefore possible to classify avalanches into two types, (i) spin rearrangements that change the total magnetization of the system, which is the bulk response and (ii) spin rearrangements that leave the magnetization unchanged. In the ferromagnetic case all avalanches were of type (i) since spins only flip from $-1$ to $+1$ and once flipped, remain in that state. Typically one is interested only in avalanches of type (i), since bulk measurements are only sensitive to them.

\begin{figure}
\includegraphics[width = 1 \columnwidth]{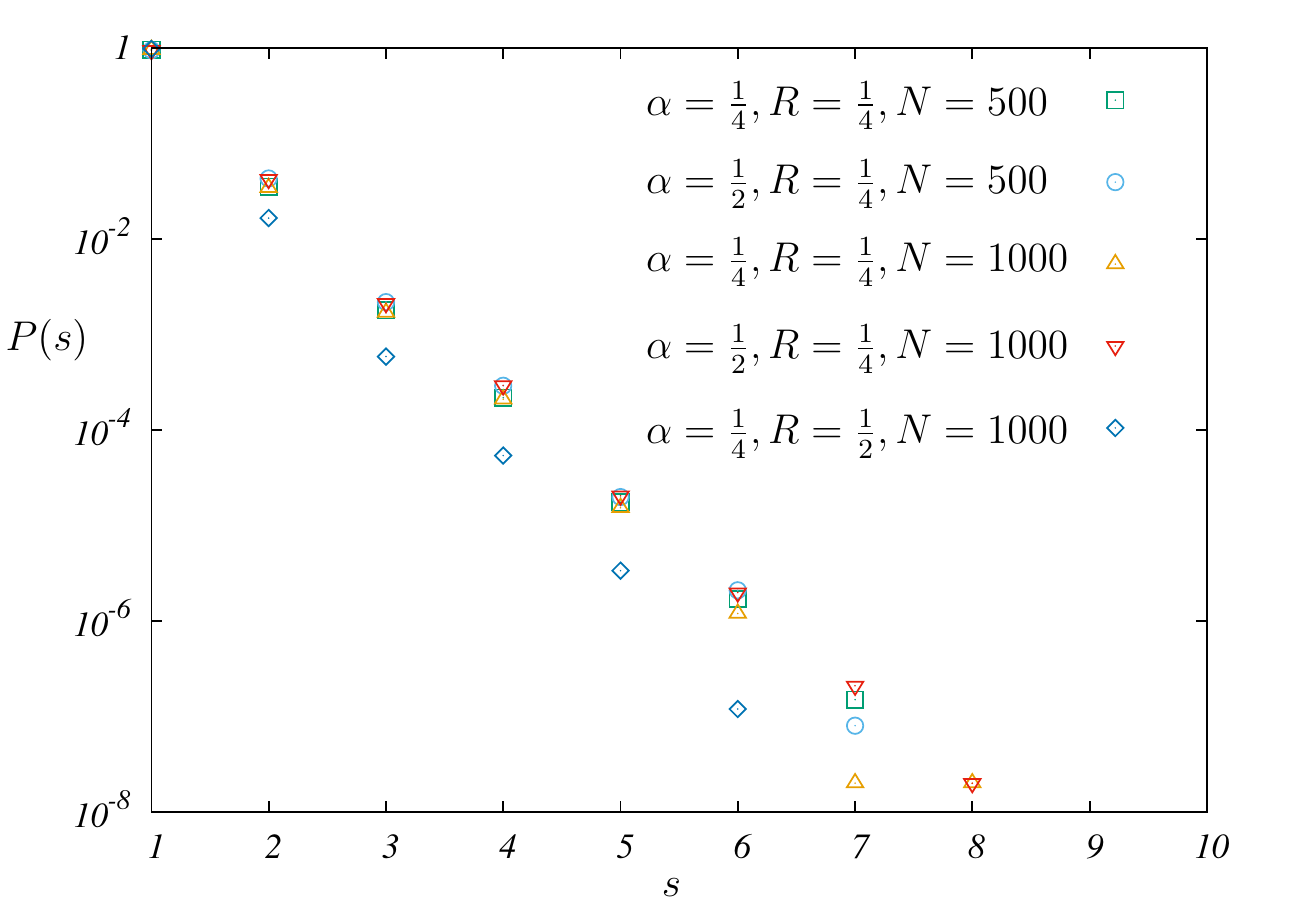}
\caption{Distribution of avalanche (changes in spin configuration) sizes $P(s)$ in the long-range antiferromagnetic RFIM for a range of model parameters. \label{fig:size_rfafim}
 The data have been averaged over $10^5$ realizations of the quenched disorder. We find that this follows a fast-decaying exponential distribution, consistent with the fact that there is no long-range order in the system.}
\end{figure}

\subsection{Absence of long-range order}
\label{absence_of_order_section}

We begin by investigating the stability of the N\'eel ground state to the presence of disorder at $H=0$ using an Imry-Ma type argument. Consider a block  of  $L$ spins in the ground state (initial configuration), numbered $k =1$ through $L$ (see Fig. \ref{fig:imry-ma_rfafim}). 
We then consider the energetic contributions from spins to the left of this block. By symmetry the spins to the right can be treated in the same manner.
The total energy contribution from the interaction of the spins in this block with all the spins to the left is denoted by $E^{l}_{\text{initial}}$. To investigate the cost of creating a domain of size $L$ in the system, we flip all the spins within the block (final configuration).
The interaction energy between the block and the left spins in this case is $E^{l}_{\text{final}}$.
We then have 
\begin{equation}
E_{l}^{\text{inital}} = -J_0\sum_{k=1}^L\sum_{n=1}^{\infty}\frac{(-1)^n}{n^{1+\alpha}},
\end{equation}
along with
\begin{equation}
E_{l}^{\text{final}} = -J_0\sum_{k=1}^{L}\left(\underbrace{\sum_{n=1}^{k-1}\frac{(-1)^n}{n^{1+\alpha}}}_\text{A}-\overbrace{\sum_{n=k}^{\infty}\frac{(-1)^n}{n^{1+\alpha}}}^\text{B}\right).
\end{equation}
In the above expression, the different terms correspond to contributions from spins to the left of the spin at site $k$, with $A$ being spins within the block and $B$ being spins outside the block. 
Next, we compute the cost of creating the domain as the energy difference between these two states. We have
\begin{equation}
\Delta E_{l} = E_{l}^{\text{final}}-E_{l}^{\text{initial}} = 2J_0\sum_{k=1}^L\sum_{n=k}^\infty \frac{(-1)^n}{n^{1+\alpha}}
\end{equation}
and as expected, we find that $\Delta E_{l} >0$. So, to examine the stability of the ordered state to disorder, we must compare this to the energy gained from disorder which scales as $\Delta E_{\text{disorder}} \sim L^{\frac{1}{2}}$. The relative contribution from these two terms in the thermodynamic limit is 
\begin{equation}
\lim_{L \to \infty} \frac{\Delta E_{l}}{\Delta E_{\text{disorder}}}= 2 J_0\lim_{L \to \infty}\underbrace{\frac{1}{L^{\frac{1}{2}}}\sum_{k=1}^L\sum_{n=k}^\infty \frac{(-1)^n}{n^{1+\alpha}}}_{I(L,\alpha)}
\end{equation}
Taking the large $L$ limit of $I(L,\alpha)$, which we do numerically, we find that 
\begin{equation}
\lim_{L\to\infty} I(L,\alpha)=0 ~~\forall{}~~ \alpha > 0.
\end{equation}
This leads to the antiferromagnetic ground state being unstable in the presence of disorder. Therefore there is no long-range order in the system at zero temperature.

\subsection{Numerical simulations}

\begin{figure}
\includegraphics[width = 1 \columnwidth]{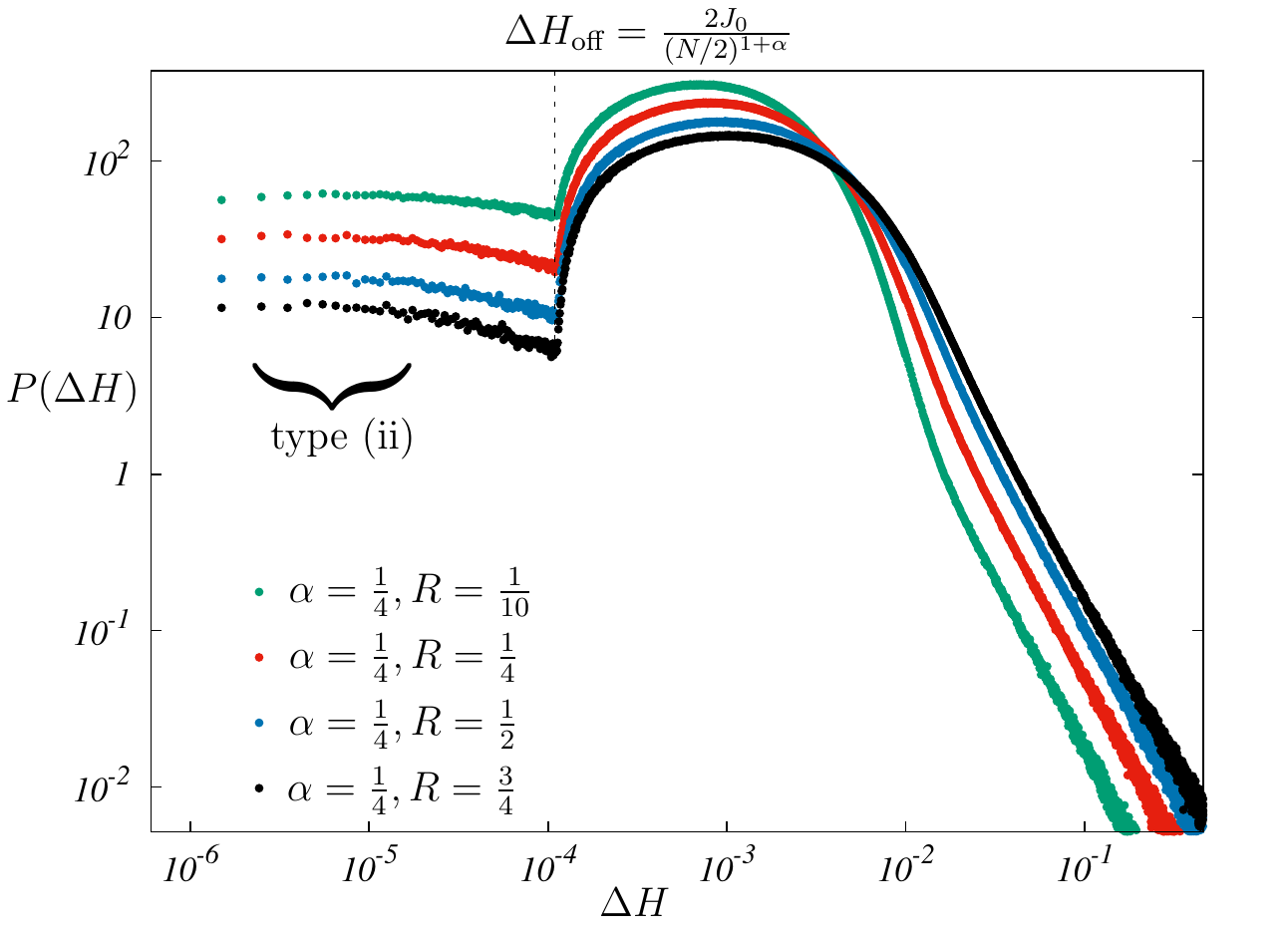}
\caption{Statistics of gaps between avalanches (changes in spin configuration) for a lattice of size $N=1000$ with $\alpha=\frac{1}{4}$ and a range of disorder strengths $R$. The plot shows two distinct types of avalanches (i) that change the total magnetization and (ii) that leave the magnetization unchanged. There is a crossover from type (ii) to type (i) dominated regions at $\Delta H_{\text{off}}$. The data have been averaged over $5\times 10^4$ realizations of the quenched disorder.}
\label{fig:gap_all}
\end{figure}

The long-ranged antiferromagnetic model does not have the useful properties of return point memory, no-crossing and Abelian dynamics that make the short-ranged ferromagnetic model analytically accessible. We therefore analyze this system using numerical simulations.
Due to the non-Abelian nature of the dynamics and the absence of return point memory, the spin configurations at a given value of $H$ depend on the details and history of the relaxation protocol. 

In our simulations we start by generating a particular realization of the quenched random field $\{h_i\}$ drawn from the exponential distribution given in Eq. (\ref{exponential_disorder}). The protocol that we employ proceeds as follows: first, we start with all spins in the $-1$ state corresponding to $H=-\infty$. We then determine the value of $H$ at which the first spin flips. This is the point at which the effective field $h_{e,i}$ at any site becomes positive, with
\begin{equation}
h_{e,i}= J_0 \sum_{\substack{j=1\\j \ne i}}^{N}\frac{S_j}{|i-j|^{1+\alpha}} + h_i + H.
\label{effective_field_RFAFIM}
\end{equation}
The system is then relaxed starting from the spin at site $1$, to obtain the configuration at that value of the field using single spin flip energy minimizing dynamics (Eq. (\ref{energy_minizing})).
Next, we compute the value of $H$ at which the next spin flips, increment $H$ to that value and relax the spins to obtain the stable configuration. This procedure is repeated until we reach the state where all spins are $+1$, which completes a `sweep' of the external field in the simulations. 
Once each configuration is stable, we measure the number of spin flips, the change in magnetization, and the gaps between successive increments in $H$. 
We collect statistics over many realizations of the quenched disorder.
All of the simulations use periodic boundary conditions and we choose units where
\begin{equation}
\sum_{i=1}^{N}\frac{J_0}{i^{1+\alpha}}=1.
\end{equation}

\begin{figure}
\includegraphics[width = 1 \columnwidth]{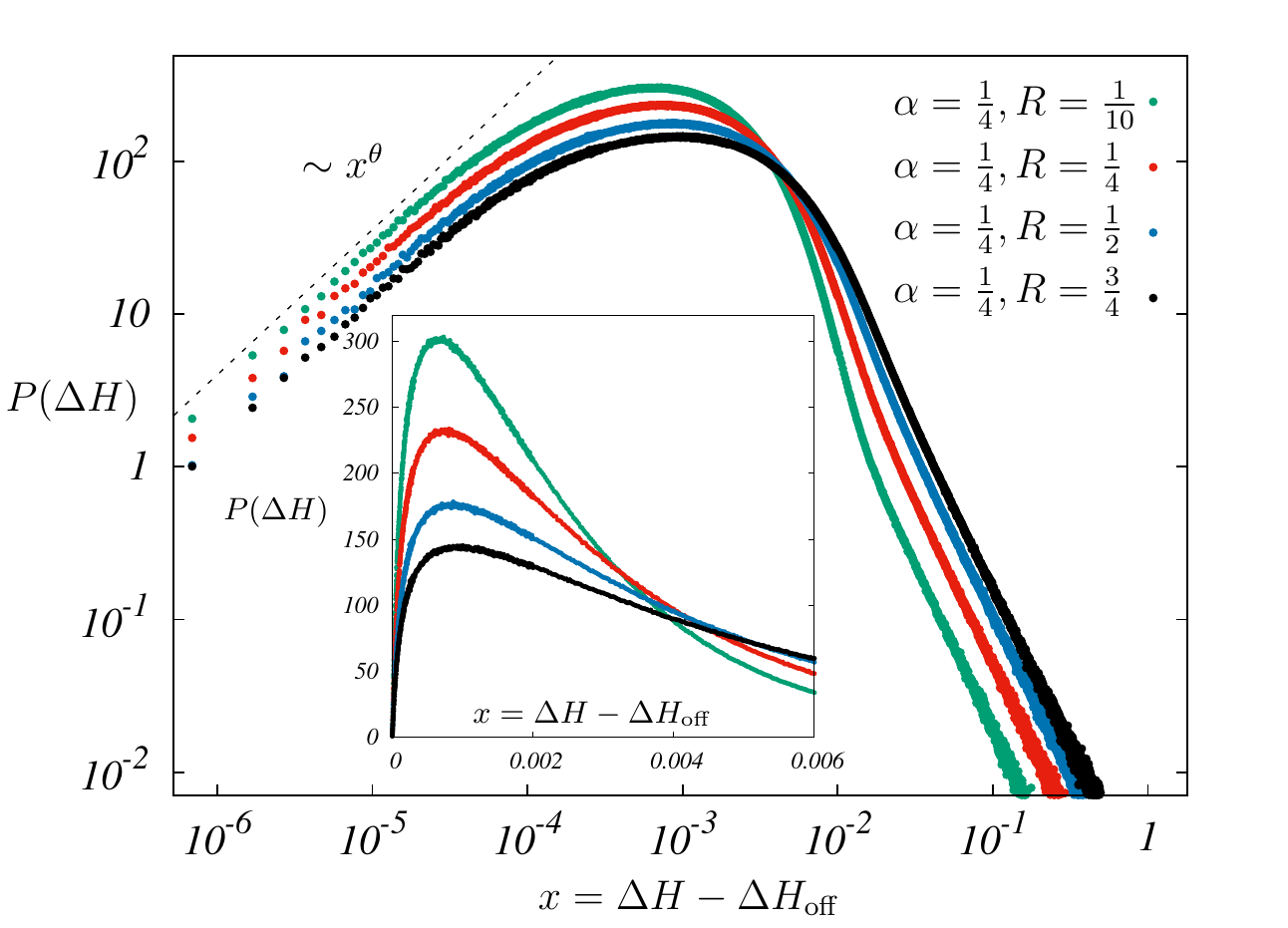}
\caption{Statistics of gaps between avalanches (changes in magnetization) for a lattice of size $N=1000$ with $\alpha=\frac{1}{4}$ and a range of disorder strengths $R$. The distribution shows a non-zero $\theta=0.95(5)$ independent of $R$ as $\Delta H \to \Delta H_{\text{off}}^+$. The data have been averaged over $5\times 10^4$ realizations of the quenched disorder. (Inset) The same data displayed in linear scale. 
\label{fig:gap_RScal}}
\end{figure}

\subsection{Statistics of avalanches}

We next examine the size and gap statistics of avalanches in this model. Since there are two types of avalanches in the system, we can study the distribution of avalanches using (a) changes in the spin configuration and (b) the jumps in magnetization. In measuring the statistics of the sizes of avalanches, we use the definition (a), which includes avalanches of types (i) and (ii). We define the size $s$ of an avalanche as the number of spins that undergo a rearrangement in an avalanche event. The distribution, $P(s)$, including avalanches of type (i) and (ii) is shown in Fig. \ref{fig:size_rfafim}, and follows an exponential distribution for the entire range of parameters that we have simulated.  This is consistent with the fact that there is no long-range ordering in the system at any finite disorder. This also indicates that both types of avalanches separately do not have any long-range ordering component.

%
%This includes avalanches that leave the overall magnetization of the system unchanged, i.e. avalanches of type (i) and type (ii) together. This distribution of avalanche sizes for a range of model parameters is shown in Fig. \ref{fig:size_rfafim}.
%We find that this distribution seems to follow an exponential distribution for the entire range of parameters that we have simulated. This is consistent with the fact that there is no long range ordering in the system.

\subsubsection*{Gap statistics}

Next, we examine the statistics of gaps between successive avalanches in this model. We can define two different gap distributions by either considering gaps between events where there is any change of spin configuration,  which includes type (i) and type (ii) avalanches, or define gaps between events that change the magnetization, which measure gaps between type (i) events.
%Once again, we can define these gaps in multiple ways. In the first case, (a) we measure the gaps between events where there is any change in the spin configuration of the system, and in the second (b) we identify the gaps between any change in the magnetization of the system. 
These two gap distributions have significantly different forms. The typical behavior of the gap distribution between {\it all} events  is shown in Fig. \ref{fig:gap_all}.
The figure clearly demarcates the two distinct populations of avalanches: the low $\Delta H$ region consists of events predominantly from type (ii) avalanches, while contributions to larger $\Delta H$ are dominated by avalanches of type (i).
The crossover from the type (ii) to type (i) dominated regions occurs at 
\begin{equation}
\Delta H_{\text{off}} = \frac{2 J_0}{\lfloor \frac{N}{2} \rfloor^{1+\alpha}},
\label{DeltaHoff}
\end{equation}
where $\lfloor \frac{N}{2} \rfloor$ is the maximum distance any spin can have from another spin on the lattice.
This can be understood in the following way: the stabilization brought about by the flip of a single spin from $-1$ to $+1$ increases the distance to failure of all the other spins by at least $\Delta H_{\text{off}}$. 
 This distance can however be decreased by avalanches with multiple spin flips (in opposite directions), as the sum of stabilizing and destabilizing effects can, for a particular spin, be made arbitrarily small. If such a spin then triggers the next avalanche, the gaps can be made arbitrarily small.
Avalanches which leave the magnetization unchanged (type (i)), can therefore be separated by gaps smaller than $\Delta H_{\text{off}}$.

\begin{figure}
\includegraphics[width = 1 \columnwidth]{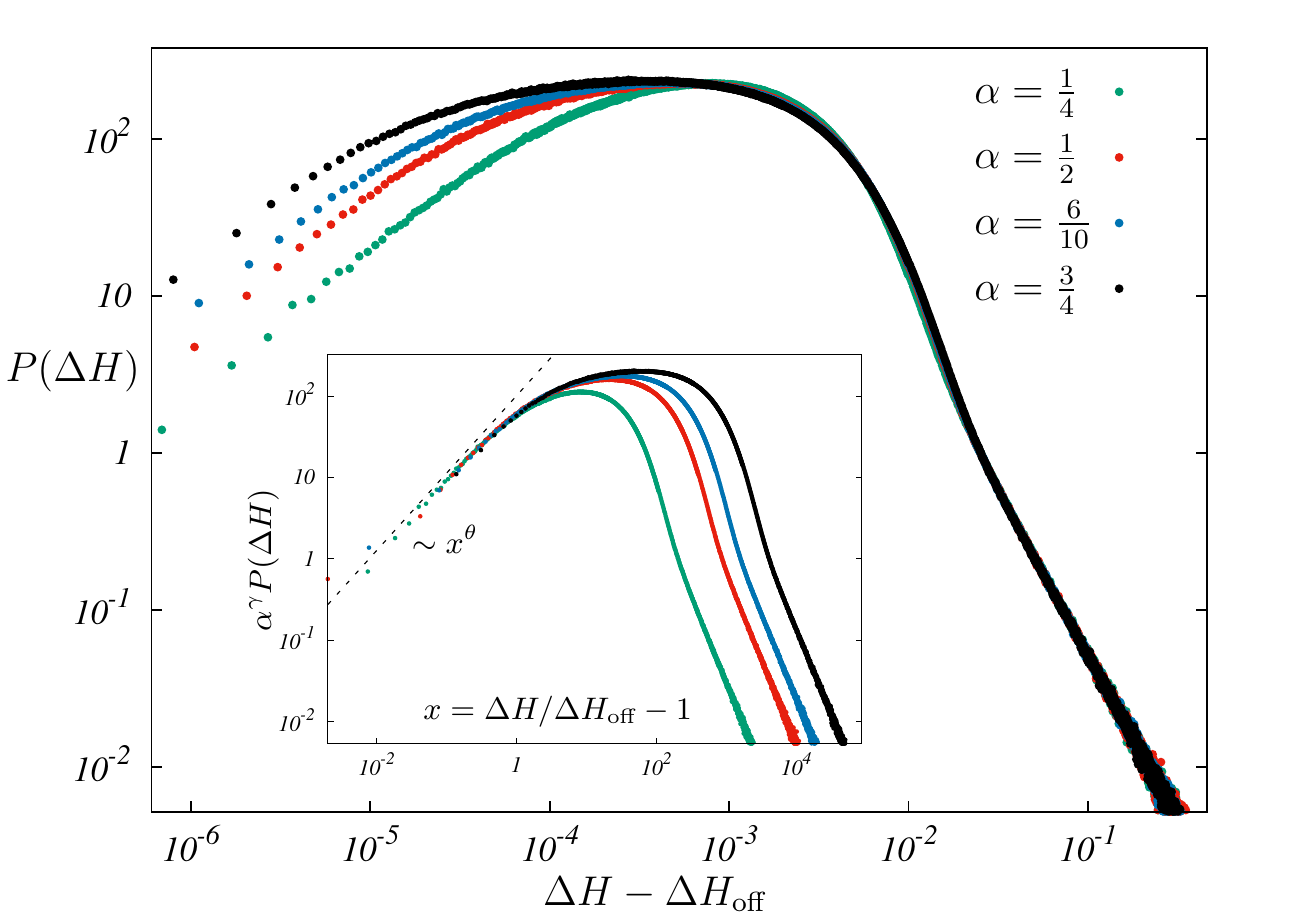}
\caption{Statistics of gaps between avalanches (changes in magnetization) for a lattice of size $N=1000$ with disorder strength $R=\frac{1}{4}$ and various interaction ranges $\alpha$. The distribution shows a non-zero $\theta=0.95(5)$ independent of $\alpha$ as $\Delta H \to \Delta H_{\text{off}}^+$. The data have been averaged over $5\times 10^4$ realizations of the quenched disorder. (Inset) The rescaled distributions show a universal behavior for small $\Delta H$, with the best fit value $\gamma = 0.50(2)$.
\label{fig:gap_alphaScal}}
\end{figure}

For avalanches that increase the magnetization, at a sufficiently large distance away from the avalanche, the effect is always of the $-1$ to $+1$ variety, since the effects from opposite spin flips cancel each other. Therefore the number of spins experiencing a stabilization smaller than $\Delta H_{\text{off}}$ scales sub-dominantly with $N$ in comparison to the number with a stabilization larger than $\Delta H_{\text{off}}$. Hence, the probability of gaps with $\Delta H < \Delta H_{\text{off}}$ also scales sub-dominantly in comparison to those with $\Delta H > \Delta H_{\text{off}}$. In contrast, events that decrease the magnetization can lead to gaps with $\Delta H < \Delta H_{\text{off}}$. However, we find from our simulations that such events are rare, and also scale sub-dominantly with $N$. In the subsequent analysis we therefore ignore gaps that succeed events that decrease the magnetization.

\begin{figure}
\includegraphics[width = 1 \columnwidth]{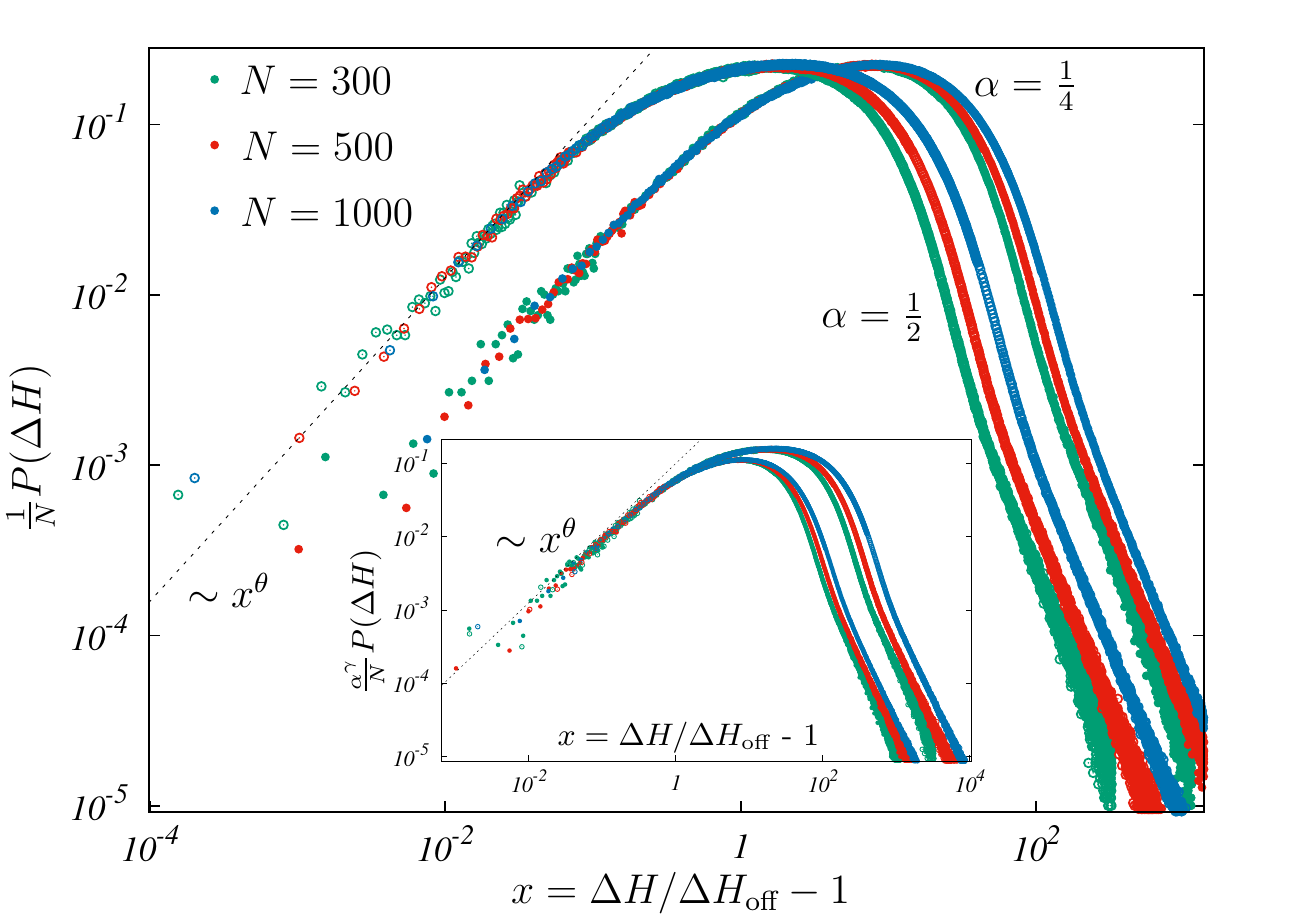}
\caption{Scaling of $P(\Delta H)$ with system size in the small $\Delta H$ regime  for $\alpha = \frac{1}{4}$ and $\frac{1}{2}$ at fixed disorder strength $R=0.25$. The distribution shows a non-zero $\theta=0.95(5)$ independent of model parameters as $\Delta H \to \Delta H_{\text{off}}^+$. The data have been averaged over $5\times 10^4$ realizations of the quenched disorder. The $\alpha = \frac{1}{2}$ plots have been shifted to the left by one decade to aid visibility. (Inset) The scaling of the same data using the scaling ansatz provided in Eq. (\ref{master_scaling_function}) with the best fit value $\gamma = 0.50(2)$ shows a very good collapse for a range of model parameters in the small $\Delta H$ regime.}
\label{fig:gap_NScal}
\end{figure}

Finally, we analyze the gap distribution between events that change the magnetization. This is the gap that one would typically measure in experiments. In this case, the region with gaps below $\Delta H_{\text{off}}$ is absent. 
%To analyze the behavior of this system near small $\Delta H$, we therefore define $\Delta H \equiv \Delta H - \Delta H_{\text{off}}$. In the thermodynamic limit, this reduces to the usual definition. 
We focus on the region close to this offset value $\Delta H \to\Delta H_{\text{off}}^+$.
We find that close to $\Delta H_{\text{off}}$, the distribution grows as a power with an non-zero $\theta$ exponent, in a range of parameters for this model. In Fig. \ref{fig:gap_RScal} we plot the distribution $P(\Delta H)$ for a range of disorder strengths $R$ at a fixed value of the range of interaction $\alpha = 1/4$. In each case we find that the distribution of gaps between avalanches has a gap up to the value $\Delta H_{\text{off}}$ and a non-trivial power law increase $P(\Delta H) \sim (\Delta H  - \Delta H_{\text{off}})^{\theta}$ for $\Delta H > \Delta H_{\text{off}}$.  The exponent $\theta$ does not seem to depend on the strength of the disorder $R$. We next analyze the nature of the gap distribution as the range of interaction is varied. In Fig. \ref{fig:gap_alphaScal} we plot this distribution for many different $\alpha$ at a fixed disorder strength $R = 1/4$. Once again we find that the distribution of gaps has a non-trivial power law increase $P(\Delta H) \sim (\Delta H  - \Delta H_{\text{off}})^{\theta}$ for $\Delta H > \Delta H_{\text{off}}$.
Since $\Delta H_{\text{off}}$ represents the smallest increment required to trigger an avalanche, the relevant scale in the small $\Delta H$ region is $\Delta H_{\text{off}}$, which depends non-trivially on $\alpha$ (Eq. (\ref{DeltaHoff})). In the inset of Fig. \ref{fig:gap_alphaScal}, we plot the distribution of gaps scaled by $\Delta H_{\text{off}}$, displaying a very good scaling collapse in the small $\Delta H$ region.
Remarkably, we find that the exponent $\theta$ does not depend on the range of interaction $\alpha$ either (we have checked this behavior up to $\alpha \le 2$).

\subsubsection*{Finite size scaling}

We next analyze the scaling properties of the gap distribution with the system size $N$. There are two relevant scales in the system, $\Delta H \sim \mathcal{O}(1/N)$ beyond which we expect successive events to occur at uncorrelated regions in space, and $\Delta H \sim \mathcal{O}(\Delta H_{\text{off}})$. As the system size is increased, $\Delta H_{\text{off}} \to 0$, and consequently the size of the gapped region also vanishes. For the region $\Delta H \gg 1/N$, we have verified the expected scaling behavior given in Eq. (\ref{scaling_form}), with avalanches occurring essentially as uncorrelated events. In the $\Delta H \sim \Delta H_{\text{off}}$ regime, the events are correlated due to the long-range interaction, controlled by the range $\alpha$, and is expected to have a different scaling with $N$.
In Fig. \ref{fig:gap_NScal} we plot $P(\Delta H)$ for various system sizes, at two different values of $\alpha = 1/4$ and $1/2$ at a fixed disorder strength $R = 1/4$. We find that once again, when the distributions are scaled by $\Delta H_{\text{off}}$, they collapse with a simple scaling with $N$. Finally, we find that for different ranges of the interaction $\alpha$ and different system sizes $N$, the distribution in the $\Delta H/\Delta H_{\text{off}} \sim \mathcal{O}(1)$ region obeys the scaling ansatz 
\begin{equation}
P(\Delta H) \sim \frac{N}{\alpha^{\gamma}} \mathcal{P}\left( \frac{\Delta H}{\Delta H_{\text{off}}} - 1 \right),
\label{master_scaling_function}
\end{equation}
with $\mathcal{P}(x) \sim x^{\theta}$ as $x \to 0^+$. Our best fit estimate is $\gamma = 0.50(2)$, and the scaling collapse using this value is illustrated in the inset of Fig. \ref{fig:gap_NScal}.

\begin{figure}
\includegraphics[width = 1 \columnwidth]{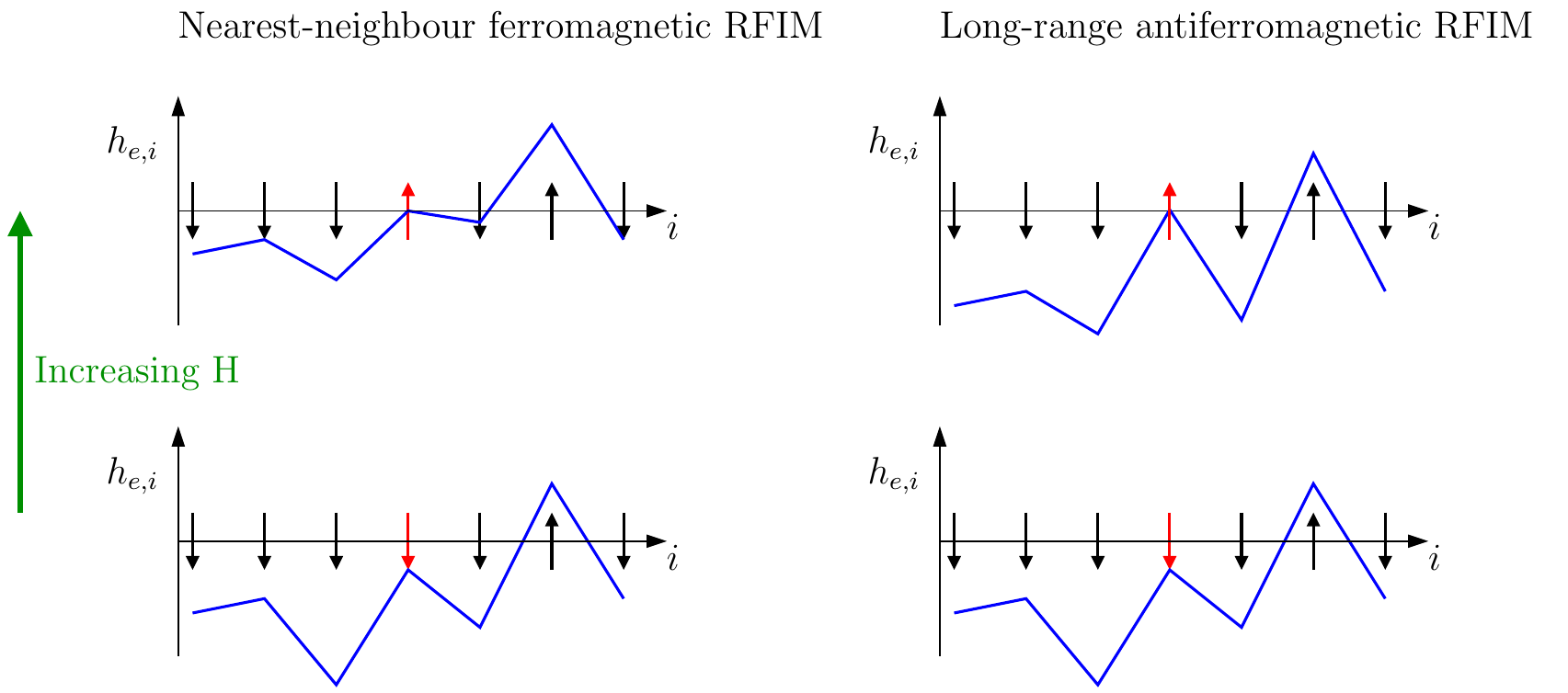}
\caption{Evolution of the local fields $h_{e,i}$ at each site, as the system undergoes an avalanche event (left) for the nearest neighbor ferromagnetic RFIM and (right) the long-range antiferromagnetic RFIM. Since the stable spin configurations are governed by $S_{i} = \text{sign}(h_{e,i})$, positive local fields correspond to spins $+1$. The smallest negative $h_{e,i}$ governs the distance (gap) to the next avalanche. In the case of the nearest neighbor ferromagnetic RFIM, since each avalanche destabilizes the neighboring spins, their ${h_{e,i}}~$s move closer to $0$. This effectively decreases the roughness of the membrane. For the long-range model, each avalanche roughens the membrane, depleting the density of near-failure regions.}
\label{interface_picture}
\end{figure}

\section{Discussion}
\label{discussion_section}

In this paper, we have examined the statistics of gaps between successive avalanches in two disordered spin models in one dimension. In the case of the nearest neighbor ferromagnetic RFIM, by mapping the avalanche events to a non-homogeneous Poisson process with a field-dependent density, we were able to relate the distribution of gaps to the underlying density of avalanche events in the system. This allowed us to derive the gap statistics exactly, and we verified our results using numerical simulations. This result confirms that the pseudo-gap exponent $\theta$ for this model is $0$ which is known from the mapping of the RFIM to the depinning process \cite{wyart_epl_2014,grinstein_prb_1983,bruinsma_prl_1984}. 
%From the study of these ferromagnetic disordered models, we have concluded that the $\theta$ exponent of the gap distribution is zero for these classes of models.

We next considered a model of Ising spins interacting via a long-range antiferromagnetic coupling, which is expected to display a non-zero $\theta$ \cite{travesset_prb_2002}. Our analysis is relevant, since models with antiferromagnetic interactions are seldom studied \cite{tadic_prl_2005,shukla_pre_2011} in relation to the avalanches that occur during a hysteresis loop.
We investigated this model using numerical simulations and analysed the features of the gap distribution. We found that this model displays a gapped behavior $P(\Delta H) = 0$ up to a system size and interaction range dependent offset value $\Delta H_{\text{off}}$ (Eq. (\ref{DeltaHoff})), and  $P(\Delta H) \sim (\Delta H - \Delta H_{\text{off}})^{\theta}$ as $\Delta H \to H_{\text{off}}^+$.
We determined $\theta \approx 0.95(5)$, independent of model parameters. An interesting property of this model is the sharp transition in $P(\Delta H)$ between regions dominated by avalanches that conserve magnetization and avalanches which change the magnetization at $\Delta H_{\text{off}}$ (see Fig. \ref{fig:gap_all}). 
%This result is interesting, since this is a microscopic Hamiltonian without any disorder in the interactions that displays this behavior, in contrast to  the well-known spin glass models \cite{wyart_anrev_2015}.

It is interesting to contrast our study of the long-range antiferromagnetic RFIM with the Coulomb glass, where avalanche statistics have been studied in detail \cite{palassini2012elementary,andresen2016charge,wyart_anrev_2015}.
%The mapping between Coulomb glass models and Ising spin models is possible at half-filling, corresponding to $M = 0$. 
The dynamics of the Coulomb glass conserve the number density, i.e. the equivalent RFIM follows a magnetization conserving Kawasaki dynamics, whereas the model studied in this paper follows a single spin flip Glauber dynamics which allows for changes in magnetization. Furthermore, avalanches in the Coulomb glass are usually studied using spatially varying (electrostatic) potentials \cite{andresen2016charge}, whereas we have used a spatially uniform external (magnetic) field. From the Coulomb glass literature, it is known that the distribution of local fields $P(h_{e,i})$ has no gap or pseudogap in one dimension (for $\alpha>0$) and a pseudogap in higher dimensions. The gap distribution $P(\Delta H)$ is directly related to the distribution of local fields $P(h_{e,i})$, and a pseudogap implies a scale free behavior of the avalanche size distribution \cite{wyart_anrev_2015}. In contrast, the model studied in this paper, displays a gap in one-dimension, and an exponentially decaying avalanche size distribution (see Fig. \ref{fig:size_rfafim}).

 %{\color{red} It is worthwhile to note the difference between the long ranged antiferromagnetic RFIM studied in this paper and another closely related model, the Coulomb glass \cite{palassini2012elementary,andresen2016charge,wyart_anrev_2015}. The gap distribution $P(\Delta H)$ is closely related to the
%distribution $P(\Delta H)$ \cite{wyart_anrev_2015} which is finite as $\Delta H \to 0$ for $\alpha > 0$ in 1D. This is in contrast with the behavior in higher
%dimensions, in which the distribution has a pseudogap with power-law
%behavior $P(h) ~ h^{d/(1+\alpha)-1}$ if $1+\alpha < d$. $h_{min}$, defined to be the smallest increment in the external field needed to trigger an avalanche, 
%scales as $N^{1/1 + \theta} << 1/N$,
%$N_e$ number of events occurring in a $\Delta H$ scales as $N^{1 + \theta}$. 
%This implies that the average avalanche in that $\Delta H$ window scales as $N^{\theta/1 + \theta}$. From these studies, it appears that in the presence of a pseudogap the avalanche size
%distribution $P(s)$ has a power-law behavior (with $<s>$ diverging in the
%thermodynamic limit), while in the absence of a pseudogap it falls
%exponentially and $<s>$ is finite.
%However, in our case we have a gap, and so $h_{min}$ scales as $N^{1+\alpha} >> 1/N$, and therefore $<s>$ is finite.}

Finally, we would like to discuss a plausible mechanism that leads to the non-trivial differences between the two models considered in this paper.
Since the stable spin configurations are governed by $S_{i} = \text{sign}(h_{e,i})$, with positive effective fields corresponding to spins $+1$, it is illuminating to parametrize the system in terms of these effective fields (Eqs. \eqref{effective_field_RFIM} and \eqref{effective_field_RFAFIM}). These values $-\infty < h_{e,i} < \infty$ can be thought of as the heights of a membrane at each site (see Fig. \ref{interface_picture}). 
As the external field $H$ is increased, the membrane drifts upwards by this amount, until an avalanche event. At each $H$, the smallest negative $h_{e,i}$ governs the distance (gap) to the next avalanche.
%We can then analyze the behvaiour of these effective fields as the system undergoes an avalanche event, and how this affects the coarse grained behavior of the membrane.

In the case of the nearest neighbor ferromagnetic RFIM, each avalanche destabilizes the neighboring spins. Since the spins only flip from $-1 \to +1$, to analyze the avalanches we only need to consider the $h_{e,i}$ which are below zero. With each avalanche event, the neighboring $h_{e,i}~$s move closer to $0$ creating a non-zero density near zero (i.e. $\theta = 0$). In higher dimensions, the Poissonian analysis of Section \ref{distribution_of_gaps_section} breaks down under certain conditions. Below a threshold disorder strength $R_c$ for $d > 2$ \cite{grinstein_prb_1983}, long-range ordering leads to a diverging average avalanche size $\langle s \rangle$, and hence correlations between values of $H$ at which avalanches occur \cite{tadic_physica_A_2000}. However, a similar argument as for one-dimension, implies  that purely destabilizing interactions lead to $\theta =0$.
%This model therefore prefers a flat interface as the ferromagnetic interaction tries to decrease its  roughness. 
%This preference for a flat membrane is what enables  mapping of this model on to the depinning model \cite{bruinsma_prl_1984,grinstein_prb_1983}. 
In the case of the long-range model, each spin flip from $-1 \to +1$ moves the local fields of its neighbors further away from $0$, roughening this membrane (see Fig. \ref{interface_picture}). Since most avalanche events are dominated by spin flips of this type, this depletes the density of events near zero, causing a gapped behavior with a non-trivial power law increase.

Alternatively, we can construct a Langevin-type equation for the evolution of $h_{e,i}$ for the long-range model. Differentiating Eq. (\ref{effective_field_RFAFIM}) with respect to the external field $H$, and using Eq. (\ref{energy_minizing}) we have 
\begin{equation}
\frac{\partial h_{e,i}}{\partial H}= \sum_{\substack{j=1\\j \ne i}}^{N}\frac{\eta_j}{|i-j|^{1+\alpha}} +1,
\label{wyart_equation}
\end{equation}
where
\begin{equation}
\eta_j=2J_0~\delta(h_{e,j})\frac{\partial h_{e,j}}{\partial H}.
\end{equation} 
Here $\eta_j$ represents a noise term that can be attributed to the quenched randomness and the interactions between spins.
This picture is then closely related to a coarse grained model that was explored by Lin \textit{et al}. \cite{wyart_epl_2014} where an evolution equation of the type Eq. (\ref{wyart_equation}) was considered, with $\eta_j$ drawn from an uncorrelated underlying distribution. In this case, it was argued, that the presence of positive as well as negative $\eta_j$, would give rise to a non-zero $\theta$.
In our case as well, $\eta_j$ can be positive or negative as the spins can flip from either $-1$ to $+1$ or vice versa with increasing external fields, providing a possible mechanism for the observed non-zero $\theta$ exponent.
However, there are crucial differences, as the noise $\eta_j$ in our model is clearly correlated. In the case of \cite{wyart_epl_2014}, the $\theta$ exponent varies with the range of interaction $\alpha$ (in some range of $\alpha$), whereas in our case this does not seem to occur. It would be interesting to explore the origin of these differences and the effects of the correlated noise in detail.

\section*{Acknowledgments}
J.N.N., K.R. and B.C. acknowledge support from NSF-DMR Grant No. 1409093 and the W. M. Keck Foundation. S.S. acknowledges support from the Indo-French Centre for the Promotion of Advanced Research (IFCPAR) under Project 5604-2.

%\clearpage
\begin{appendix}
\label{appendix}

\section{Joint distribution of successive avalanches}
\label{joint_distribution_section}

In this Appendix we analyze the joint density of successive avalanches in the RFIM with ferromagnetic coupling of range $\delta$, with a quenched random field $\{h_i\}$ at each site. 
%Every configuration of quenched random fields $\{ h_i \}$, is drawn from a mutually independent underlying distribution $\phi(h,R)$. 
As argued in Section \ref{distribution_of_gaps_section}, the spins in the system can be grouped into clusters that undergo avalanches together as the external field $H$ is increased monotonically and quasi-statically.
Corresponding to each realization of $\{h_i\}$, we have a unique cluster decomposition $\{ c_j, H_j \}$, with the spins being grouped into clusters $c_j = \{S_{j,1}, S_{j,2} ...\}$ with $j = 1,2...N_a$ and $N_a$ is the total number of avalanches in the realization. The $\{ H_i \}$ correspond to the values at which each cluster undergoes an avalanche. This set varies for each realization, and we are interested in the statistics of the ordered set $\{H_1 < H_2 ... < H_{N_a}\}$. The disorder average can now be performed in two steps, first over all realizations of the quenched randomness consistent with a cluster decomposition, and then over all possible cluster decompositions
\begin{equation}
\langle ...\rangle_{\{ h_i \}} = \langle \langle  ...\rangle_{\{h_{i} \} |\{ c_j\}} \rangle_{\{c_j\}}.
\end{equation}
We next consider the joint density  $\rho(H,H'|\{c_j\})$ such that, given a cluster decomposition $\{c_j\}$, two successive avalanches occur at values $H$ and $H'$.
We have
\begin{equation}
\rho(H,H'|\{c_j\}) = \langle \rho(H,H'| \{c_j\},\{h_i\} ) \rangle_{\{h_{i} \}| c_j },
\label{first_step_disorder}
\end{equation}
along with
\begin{equation}
\rho(H,H') = \langle \rho(H,H'| \{c_j\} ) \rangle_{\{c_j\}} .
\label{second_step_disorder}
\end{equation}
where $\rho(H,H')$ is the probability that two successive avalanches occur at $H$ and $H'$ over all realizations of disorder. We first consider the disorder average in Eq. (\ref{first_step_disorder}). We define $\rho_\text{ind}(H,H'|\{c_j\})$ as the two point density of successive avalanches computed using the one point density $\rho(H|\{c_j\})$ and assuming that events at $H$ and $H'$ are independent. We are interested in the correlation between the events at $H$ and $H'$ which can be estimated by the deviation from $\rho_\text{ind}(H,H'|\{c_j\})$. 
Since we are only concerned with successive events, the contribution to
this deviation $\Delta_j$ occurs only through the interaction of the avalanche at $H$ with its neighboring clusters. Next, since the clusters interact through their boundaries, this deviation can be expected to scale as
\begin{equation}
\Delta_j=|\rho_\text{ind}(H,H'|\{c_j\})-\rho(H,H'|\{c_j\})|  \propto \delta \langle s\rangle_j^{\frac{d-1}{d}}.
\end{equation}
However, the number of clusters unaffected by this avalanche scales as $N_a=N/\langle s \rangle_j$ where $\langle s\rangle_j$ is the average cluster size in $\{c_j\}$. Now, since $\rho_\text{ind}(H,H'|\{c_j\})$ has contributions from all the clusters in the system, the relative importance of correlations therefore scales as $\delta \langle s \rangle_j^{\frac{2d-1}{d}}/N$.

We can now estimate the importance of correlations over all realizations of the disorder as
\begin{equation}
\Delta= \langle\Delta_j\rangle_{c_j}  \propto \delta \langle \langle s\rangle_j^{\frac{d-1}{d}} \rangle_{\{ c_j \}}.
\end{equation}
In the absence of long-range ordering $\langle s\rangle_j$ has a well-defined distribution with no diverging moments, and therefore $\langle \langle s\rangle_j^{\mu} \rangle_{\{ c_j \}} \sim \langle s \rangle^\mu$.
The relative number of correlated events therefore scales as $\delta \langle s \rangle^{\frac{2d-1}{d}}/N$ and vanishes in the thermodynamic limit as long as $\langle s \rangle$ and $\delta$ remain finite. In our system, the interaction is finite ranged and therefore $\delta$ is finite. In addition, there is no long-range ordering in the system, hence $ \langle s \rangle$ is finite. We can therefore treat the avalanches as independent events in the thermodynamic limit.

\section{Distribution of gaps in the non-homogeneous Poisson process}
\label{gap_distribution_appendix}

In this appendix we derive an expression for the distribution of gaps between avalanches in a finite window of the external field. To do this we consider a system starting at $H_1$ and compute the distribution of gaps between events as the field is increased up to $H_2 > H_1$. We first compute the cumulative probability $S(x;H_1,H_2)$ of the occurrence of a gap of size $\Delta H > x$ over the magnetization sweep from $H_1$ to $H_2$. This can be computed as the probability that an event occurs at a value $H' + x$ with no events between $H'$ and $H' + x$, with $H_1 < H' < H_2 - x$. This ensures that the avalanche is preceded by a gap of a size at least $x$. We then have 
\begin{equation}
S(x;H_1,H_2) = \int_{H_1}^{H_2 - x} \frac{\rho (H' + x)}{N(H_1,H_2)} e^{-\int_{H'}^{H'+x} \rho(y) dy }  dH',
\label{integrated_pdf}
\end{equation}
where $N(H_1,H_2) = \int_{H_1}^{H_2} \rho(H') dH'$ is the expected number of events in the interval $[H_1,H_2]$. 
The gap distribution is then simply 
\begin{equation}
P(x;H_1,H_2) = -\frac{dS(x;H_1,H_2)}{dx}.
\end{equation}
Taking the derivative of Eq (\ref{integrated_pdf}) and integrating by parts, we arrive at \cite{yakovlev_arxiv_2005,shcherbakov_prl_2005}

\begin{eqnarray}
\nonumber
&&P(\Delta H;H_1,H_2) =\\
\nonumber
&&\int_{H_1}^{H_2 - \Delta H} \frac{ \rho(H')\rho (H'+\Delta H)}{N(H_1,H_2)} e^{-\int_{H'}^{H'+\Delta H} \rho(y) dy } d H'\\
\nonumber
&&~~~~~~+ \frac{\rho(H_1 + \Delta H) e^{
-\int_{H_1}^{H_1 + \Delta H} \rho(y) dy }}{N(H_1,H_2)} .
\label{window_distribution}
\end{eqnarray}
The distribution of gaps over the entire sweep from $-\infty$ to $+\infty$ can then be derived using
\begin{equation}
P(\Delta H) = \lim_{H_1 \to -\infty} \lim_{H_2 \to +\infty} P(\Delta H;H_1,H_2).
\end{equation}
Since $\rho(H) \to 0$ as $H \to \pm \infty$, this yields Eq. (\ref{full_sweep_pdf}).

\section{Generating function for avalanche sizes: $G(x,H)$}
\label{generating_function_appendix}

\begin{figure}
\centering
\includegraphics[width = 1 \columnwidth]{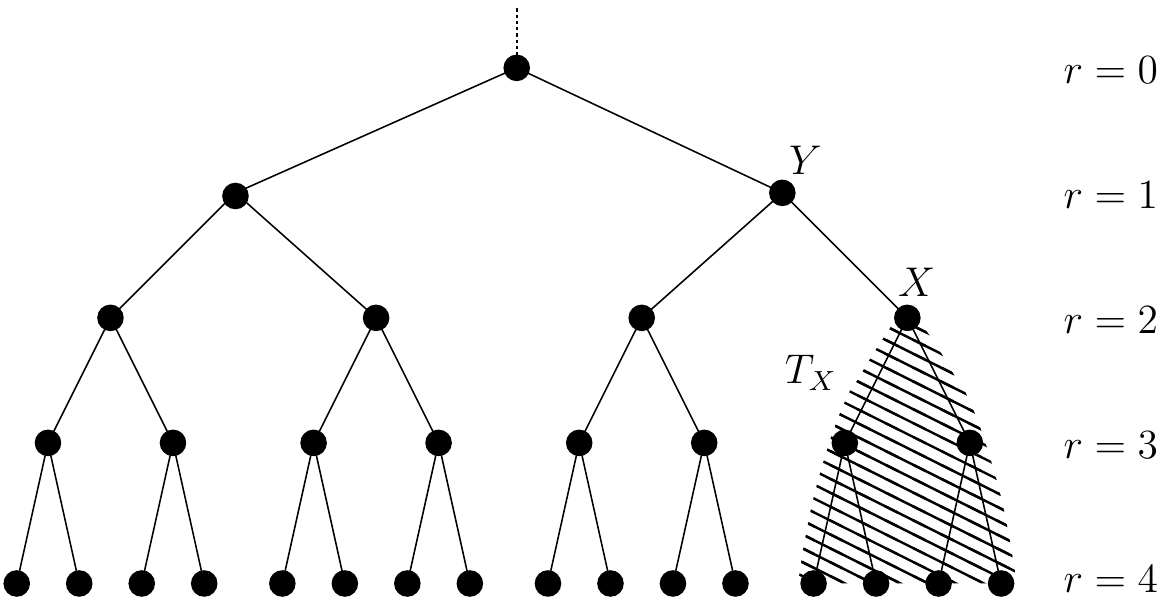}
\caption{Illustration of a Bethe lattice with coordination number $z = 3$, along with two of its associated Cayley subtrees with four generations ($r = 4$). The site at $r = 0$ is the origin. The shaded region denotes a subtree $T_X$ rooted at $X$ with a parent node $Y$. The linear chain, considered in this paper, corresponds to $z=2$.}
\label{fig:subtree}
\end{figure}

In this appendix, we compute the generating function for the avalanche size distribution $G(x,H)$ for the nearest-neighbor ferromagnetic RFIM on a Bethe lattice with coordination number $z$ at zero temperature, reproducing the work of Sabhapandit \textit{et al}. \cite{sanjib_thesis,sanjib_jstatphys_2000}. The case $z = 2$ reduces to the one dimensional model considered in Section \ref{nearest_neighbour_rfim}. The Hamiltonian of the system is given by $\mathcal{H}=-J \sum_{\langle i,j \rangle} S_i S_j - \sum_i \left(h_i+H\right) S_i$, where $\langle \rangle$ denotes nearest neighbors on the Bethe lattice (see Fig. \ref{fig:subtree}).

\subsection{Magnetization per site}

We begin with the system at $H=-\infty$ (i.e. all the spins $S_i = -1$) and increase the external field to a value $H > -\infty$. Due to the return point memory of this model, the resulting configuration is exactly the same for any history of external field increments. We can therefore directly increase the field from $-\infty$ to $H$. 
We define $p_m \equiv p(m,H)$ as the probability that a spin $S_i$ is $+1$ at $H$ {\it given} that $m$ of its neighbors are $+1$. This is given by the probability that the local field $h_{e,i}$ at this site is positive. This can be computed as
\begin{equation}
p(m,H)=P(h_{e,i}>0)=\int_{J(z-2m)-H}^\infty\phi(h)dh.
\label{eq:pm}
\end{equation}

Due to the Abelian property the final stable configuration is independent of the order in which the spins are relaxed. We therefore choose a relaxation protocol that propagates upwards from the last generation of the Bethe lattice (see Fig. \ref{fig:subtree}). We define $P^{(r)}(H)$ as the probability that a spin in the $r^{th}$ generation is $+1$ when its parent spin at $(r-1)$ is $-1$, with all its descendants in their stable configuration. We then have 
%From equation \eqref{eq:pm} we can compute $P^{(r)}(H)$, if we know the probability of $m$ of the descendants being up which is given by $\binom{z-1}{m} \left(P^{(r+1)}(H)\right)^{m}\left(1-P^{(r+1)}(H)\right)^{z-1-m}$. So we get
\begin{align}
\label{eq:pr_bethe}
P^{(r)} (H)=\sum_{m=0}^{z-1}&\binom{z-1}{m}\left(P^{(r+1)}(H)\right)^{m} \\ \nonumber
&\cdot \left(1-P^{(r+1)}(H)\right)^{z-1-m}p(m,H)
.
\end{align}

Since the sites deep in the tree are all equivalent, $P^{(r)}(H) \rightarrow P^*(H)$ for `$r$' deep inside the tree. The value of $P^* \equiv P^*(H)$ can therefore be computed by substituting this into Eq. \eqref{eq:pr_bethe}, yielding
\begin{align}
\label{eq:pst}
P^{*} (H)=\sum_{m=0}^{z-1}&\binom{z-1}{m}\left(P^{*}(H)\right)^{m} \\ \nonumber
& \cdot \left(1-P^{*}(H)\right)^{z-1-m}p(m,H).
\end{align}
Choosing the site in the bulk as the origin, i.e. $r = 0$, the magnetization per site can be computed by evaluating $P^{(r = 0)}(H)$. This is simply the probability that the spin at the origin is $+1$. We have
\begin{align}
P^{(0)} (H)=\sum_{m=0}^{z}&\binom{z}{m}\left(P^{*}(H)\right)^{m}\\ \nonumber 
&\cdot \left(1-P^{*}(H)\right)^{z-m}p(m,H).
\end{align}

The magnetization per site of the nearest neighbor ferromagnetic RFIM on the Bethe lattice is therefore determined by the behavior of $P^*(H)$. From Eq. \eqref{eq:pst}, it can be seen that the equation determining $P^*(H)$ is of degree $z-1$. Therefore, for the linear chain (the $z=2$ Bethe lattice), this equation is linear, leading to a magnetization that is a continuous function of the external field $H$.

\subsection{Avalanche size distribution}

Next, consider the Cayley tree rooted at some spin $X$ at generation `$r$' deep in the Bethe lattice (see Fig. \ref{fig:subtree}). The subtree formed by $X$ and all its descendants is referred to as the subtree rooted at $X$ and denoted by $T_X$. Let $Q_n$ be the probability that exactly `$n$' spins in $T_X$ that were $-1$ when the parent spin at $(r-1)$ was $-1$, flip to $+1$ when the parent spin flips to $+1$. If the spin at $X$ was already $+1$, which occurs with the probability $P^*$, the spins in the subtree would be unaffected by the flip of the spin at $Y$ and we obtain
\begin{equation}
P^*(H)+\sum_{n=0}^{\infty}Q_n(H)=1.
\end{equation}
By definition, $Q_0$ is the probability that the spin at $X$ was $-1$ when $Y$ was $-1$, and remained $-1$ when $Y$ flipped to $+1$. The probability of any descendant of $X$ being $+1$ when $X$ is $-1$ is given by $P^*$, hence the probability that $m$ of the descendants of $X$ were $+1$ after the relaxation is given by $\binom{z-1}{m}(P^*)^m (1-P^*)^{z-1-m}$. Now, if $m$ of its descendants were $+1$, the probability that $X$ remains $-1$ after the spin flip at $Y$ is $(1-p_{m+1})$. We then have
\begin{equation}
Q_0=\sum_{m=0}^{z-1}\binom{z-1}{m}(P^*)^{m}(1-P^*)^{(z-1-m)}(1-p_{m+1}).
\end{equation}

\begin{figure}
  \centering
  \includegraphics[width = 1 \columnwidth]{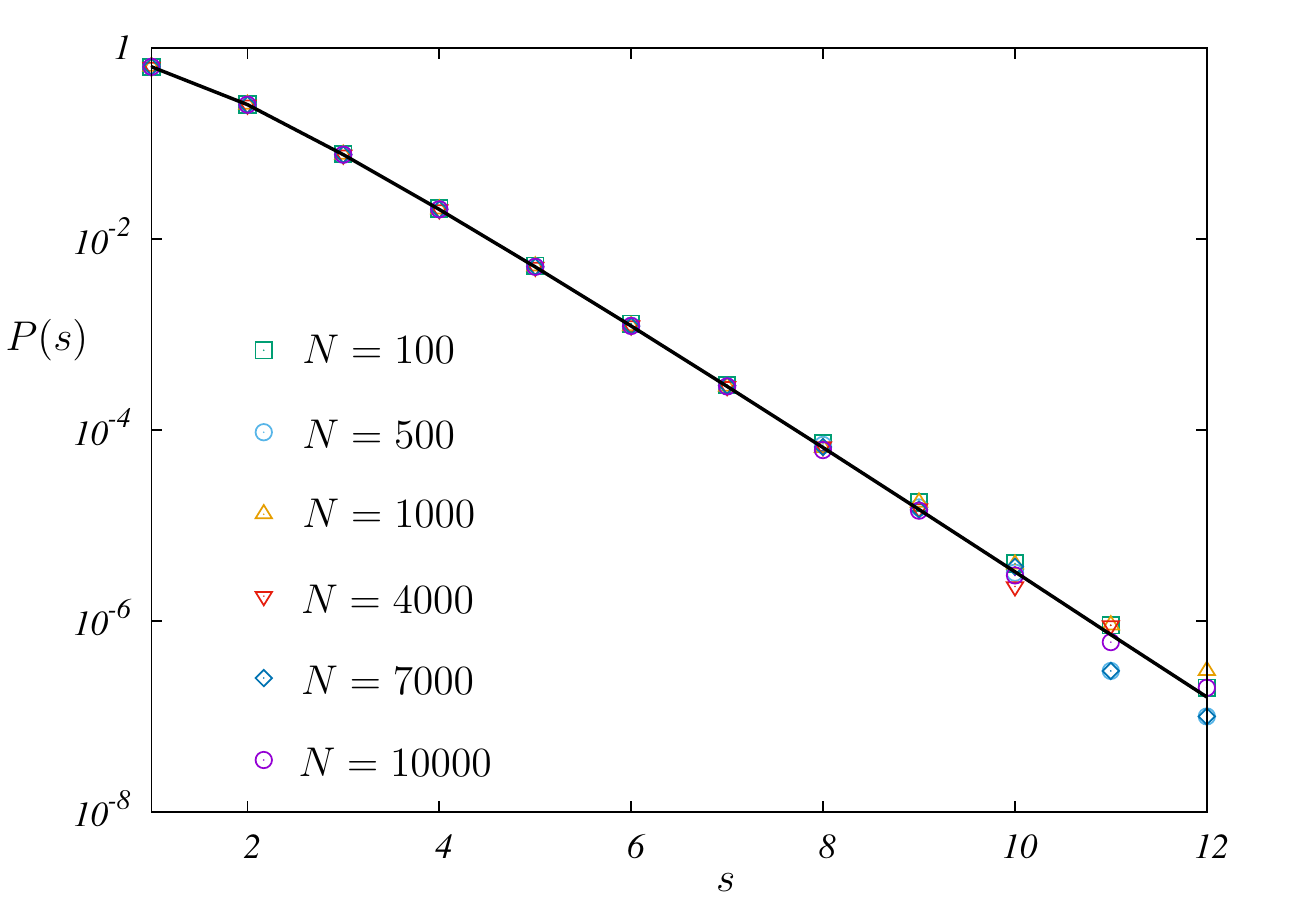}
\caption{Size distribution of avalanches for the one dimensional RFIM with random fields drawn from a uniform distribution (Eq. \eqref{uniform_disorder}) with $R=5$ at an external field $H=1$, for different system sizes $N$. The data have been averaged over $10^7$ realizations of the disorder. The bold line corresponds to the analytic expression for $N \to \infty$ computed using Eq. \eqref{eq:gx}.}
\label{uniform_avalanche_fig}
\end{figure}

Now, we can recursively compute $Q_n$ for general $n$. For example $Q_1$ is the probability that the spin at $X$ which was $-1$ when $Y$ was $-1$ flipped to $+1$ when $Y$ flipped to $+1$ and among the $z-1-m$ descendants of $X$ which were $-1$, none of them flipped to $+1$ when $X$ flipped. This occurs with a probability of $(p_{m+1}-p_m)Q_0^{z-1-m}\left(P^*\right)^m$. So we have
\begin{equation}
Q_1=\sum_{m=0}^{z-1}\binom{z-1}{m}Q_0^{z-1-m}\left(P^*\right)^m(p_{m+1}-p_m).
\end{equation}
We can similarly compute $Q_n$ recursively for higher $n$, noting the fact that determining $Q_n$ requires only the knowledge of $Q_i \ \forall\ i<n$. The recursion is given by
\begin{align*}
Q_n=\sum_{m=0}^{z-1}&\binom{z-1}{m}\left(P^*\right)^m(p_{m+1}-p_m) \\
&\cdot\left[\sum_{\{n_i\}=0}^{\infty}\left(\prod_{i=1}^{z-1-m}Q_{n_i}\right)\delta\left(\sum n_i,~n-1\right)\right],
\end{align*}
where $\delta$ represents the Kronecker delta function.
The recursion relation becomes much simpler when we express it in terms of the generating function $Q(x)=\sum_{n=0}^\infty Q_n x^n$. We have 
\begin{align}
Q(x)=Q_0+x\sum_{m=0}^{z-1}&\binom{z-1}{m}\left(P^*\right)^m Q(x)^{z-1-m}\\
&\cdot (p_{m+1}-p_m).
\label{eq:qx}
\end{align}

\begin{figure}
  \centering
  \includegraphics[width = 1 \columnwidth]{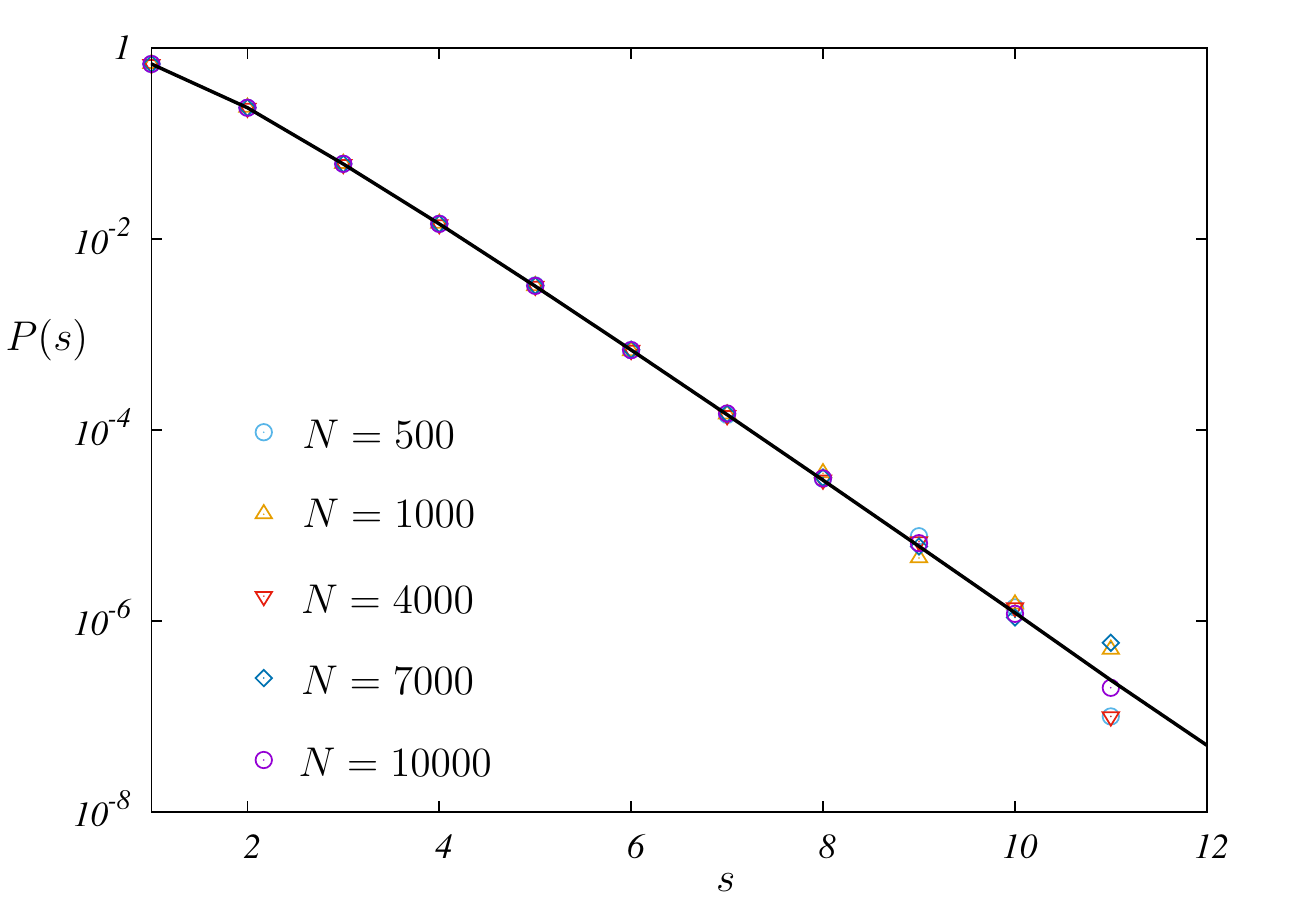}
  \caption{Size distribution of avalanches for the one dimensional RFIM with random fields drawn from an exponential distribution (Eq. \eqref{exponential_disorder}) with $R=5$ at an external field $H=1$, for different system sizes $N$. The data have been averaged over $10^7$ realizations of the disorder. The bold line corresponds to the analytic expression for $N \to \infty$ computed using Eq. \eqref{eq:gx}.}
  \label{exponential_avalanche_fig}
\end{figure}

Next, we define $P(s) \equiv P(s,H)$ as the probability that an avalanche of size $s$ is initiated at the origin when the field is increased from $H$ to $H+dH$. $P(1,H)dH$ is the probability that an avalanche of size $1$ is initiated at the origin when the field is increased from $H$ to $H+dH$ i.e. no descendant spin which was $-1$ flipped in response to this avalanche at the origin. If $m$ of the descendants of the origin were $+1$ at $H$, the probability of the spin at the origin flipping from $-1$ to $+1$ during the field increment $H \to H+dH$ is the probability that the local disorder field $(h_0)$ satisfies $J(2m-z)+H+h_0<0$ {\it and} $J(2m-z)+H+dH+h_0>0$. This is simply given by $\phi(J(z-2m)-H)dH$. We therefore have
\begin{equation}
P(1,H)=\sum_{m=0}^{z}\binom{z}{m}\left(P^*\right)^m Q_0^{z-m} \phi(J(z-2m)-H).
\end{equation}

Following the same arguments as for $Q_n$, we can recursively compute $P(s,H)$ and then express it in terms of the generating function $G(x,H)=\sum_{s=1}^\infty P(s,H) x^s$. We have
\begin{equation}
  G(x,H)=x\sum_{m=0}^{z}\left(P^*\right)^m \left(Q(x)\right)^{z-m}\phi(J(z-2m)-H).
  \label{eq:gx}
  \end{equation}

We then compute this generating function for avalanche sizes $G(x,H)$ for the RFIM in one dimension (the $z=2$ Bethe lattice) with uniform and exponential disorder distributions chosen from Eqs. \eqref{uniform_disorder} and \eqref{exponential_disorder}. In both cases, we compare the result obtained to the distributions obtained by direct numerical simulations (see Figs. \ref{uniform_avalanche_fig} and \ref{exponential_avalanche_fig}). We find that the avalanche size distribution is a fast-decaying exponential in the region of parameters that we explore and the simulation results agree well with the analytical results. 

Finally, we use the expression in Eq. \eqref{eq:gx} to compute the quantity $G(1,H)$ which is the average density of avalanche events at a given $H$ as $\rho(H) = N G(1,H)$ (Eq. \eqref{eq:rho_to_G}). Using the expressions in Eqs. \eqref{eq:pst}, \eqref{eq:qx} and \eqref{eq:gx}, along with the disorder distributions given in Eq. \eqref{exponential_disorder} and \eqref{uniform_disorder}, we arrive at the expressions announced in Eq. \eqref{nn_uniform_g1} and Eq. \eqref{nn_exponential_g1}.
 
\end{appendix}

\bibliography{bib} 

\begin{thebibliography}{34}%
\makeatletter
\providecommand \@ifxundefined [1]{%
 \@ifx{#1\undefined}
}%
\providecommand \@ifnum [1]{%
 \ifnum #1\expandafter \@firstoftwo
 \else \expandafter \@secondoftwo
 \fi
}%
\providecommand \@ifx [1]{%
 \ifx #1\expandafter \@firstoftwo
 \else \expandafter \@secondoftwo
 \fi
}%
\providecommand \natexlab [1]{#1}%
\providecommand \enquote  [1]{``#1''}%
\providecommand \bibnamefont  [1]{#1}%
\providecommand \bibfnamefont [1]{#1}%
\providecommand \citenamefont [1]{#1}%
\providecommand \href@noop [0]{\@secondoftwo}%
\providecommand \href [0]{\begingroup \@sanitize@url \@href}%
\providecommand \@href[1]{\@@startlink{#1}\@@href}%
\providecommand \@@href[1]{\endgroup#1\@@endlink}%
\providecommand \@sanitize@url [0]{\catcode `\\12\catcode `\$12\catcode
  `\&12\catcode `\#12\catcode `\^12\catcode `\_12\catcode `\%12\relax}%
\providecommand \@@startlink[1]{}%
\providecommand \@@endlink[0]{}%
\providecommand \url  [0]{\begingroup\@sanitize@url \@url }%
\providecommand \@url [1]{\endgroup\@href {#1}{\urlprefix }}%
\providecommand \urlprefix  [0]{URL }%
\providecommand \Eprint [0]{\href }%
\providecommand \doibase [0]{http://dx.doi.org/}%
\providecommand \selectlanguage [0]{\@gobble}%
\providecommand \bibinfo  [0]{\@secondoftwo}%
\providecommand \bibfield  [0]{\@secondoftwo}%
\providecommand \translation [1]{[#1]}%
\providecommand \BibitemOpen [0]{}%
\providecommand \bibitemStop [0]{}%
\providecommand \bibitemNoStop [0]{.\EOS\space}%
\providecommand \EOS [0]{\spacefactor3000\relax}%
\providecommand \BibitemShut  [1]{\csname bibitem#1\endcsname}%
\let\auto@bib@innerbib\@empty
%</preamble>
\bibitem [{\citenamefont {Sethna}\ \emph {et~al.}(2006)\citenamefont {Sethna},
  \citenamefont {Dahmen},\ and\ \citenamefont {Perkovic}}]{sethna_review_2004}%
  \BibitemOpen
  \bibfield  {author} {\bibinfo {author} {\bibfnamefont {J.~P.}\ \bibnamefont
  {Sethna}}, \bibinfo {author} {\bibfnamefont {K.~A.}\ \bibnamefont {Dahmen}},
  \ and\ \bibinfo {author} {\bibfnamefont {O.}~\bibnamefont {Perkovic}},\ }in\
  \href {\doibase https://doi.org/10.1016/B978-012480874-4/50013-0} {\emph
  {\bibinfo {booktitle} {The Science of Hysteresis}}},\ \bibinfo {editor}
  {edited by\ \bibinfo {editor} {\bibfnamefont {G.}~\bibnamefont {Bertotti}}\
  and\ \bibinfo {editor} {\bibfnamefont {I.~D.}\ \bibnamefont {Mayergoyz}}}\
  (\bibinfo  {publisher} {Academic Press},\ \bibinfo {address} {Oxford},\
  \bibinfo {year} {2006})\ pp.\ \bibinfo {pages} {107 -- 179}\BibitemShut
  {NoStop}%
\bibitem [{\citenamefont {Sethna}\ \emph {et~al.}(2001)\citenamefont {Sethna},
  \citenamefont {Dahmen},\ and\ \citenamefont
  {Myers}}]{sethna_crackling_review_nature_2001}%
  \BibitemOpen
  \bibfield  {author} {\bibinfo {author} {\bibfnamefont {J.~P.}\ \bibnamefont
  {Sethna}}, \bibinfo {author} {\bibfnamefont {K.~A.}\ \bibnamefont {Dahmen}},
  \ and\ \bibinfo {author} {\bibfnamefont {C.~R.}\ \bibnamefont {Myers}},\
  }\href@noop {} {\bibfield  {journal} {\bibinfo  {journal} {Nature}\ }\textbf
  {\bibinfo {volume} {410}},\ \bibinfo {pages} {242} (\bibinfo {year}
  {2001})}\BibitemShut {NoStop}%
\bibitem [{\citenamefont {Barkhausen}(1919)}]{barkhausen_zphys_1919}%
  \BibitemOpen
  \bibfield  {author} {\bibinfo {author} {\bibfnamefont {H.}~\bibnamefont
  {Barkhausen}},\ }\href@noop {} {\bibfield  {journal} {\bibinfo  {journal} {Z.
  Phys.}\ }\textbf {\bibinfo {volume} {20, 401.}} (\bibinfo {year}
  {1919})}\BibitemShut {NoStop}%
\bibitem [{\citenamefont {Gutenberg}\ and\ \citenamefont
  {Richter}(1944)}]{gutenberg_bul.seis_1944}%
  \BibitemOpen
  \bibfield  {author} {\bibinfo {author} {\bibfnamefont {B.}~\bibnamefont
  {Gutenberg}}\ and\ \bibinfo {author} {\bibfnamefont {C.~F.}\ \bibnamefont
  {Richter}},\ }\href@noop {} {\bibfield  {journal} {\bibinfo  {journal} {Bull.
  Seismol. Soc. Am.}\ }\textbf {\bibinfo {volume} {34}},\ \bibinfo {pages}
  {185} (\bibinfo {year} {1944})}\BibitemShut {NoStop}%
\bibitem [{\citenamefont {Gutenberg}\ and\ \citenamefont
  {Richter}(1956)}]{gutenberg_bul.seis_1956}%
  \BibitemOpen
  \bibfield  {author} {\bibinfo {author} {\bibfnamefont {B.}~\bibnamefont
  {Gutenberg}}\ and\ \bibinfo {author} {\bibfnamefont {C.~F.}\ \bibnamefont
  {Richter}},\ }\href@noop {} {\bibfield  {journal} {\bibinfo  {journal} {Ann.
  Geophys.}\ }\textbf {\bibinfo {volume} {9}},\ \bibinfo {pages} {1} (\bibinfo
  {year} {1956})}\BibitemShut {NoStop}%
\bibitem [{\citenamefont {Shcherbakov}\ \emph {et~al.}(2005)\citenamefont
  {Shcherbakov}, \citenamefont {Yakovlev}, \citenamefont {Turcotte},\ and\
  \citenamefont {Rundle}}]{shcherbakov_prl_2005}%
  \BibitemOpen
  \bibfield  {author} {\bibinfo {author} {\bibfnamefont {R.}~\bibnamefont
  {Shcherbakov}}, \bibinfo {author} {\bibfnamefont {G.}~\bibnamefont
  {Yakovlev}}, \bibinfo {author} {\bibfnamefont {D.~L.}\ \bibnamefont
  {Turcotte}}, \ and\ \bibinfo {author} {\bibfnamefont {J.~B.}\ \bibnamefont
  {Rundle}},\ }\href@noop {} {\bibfield  {journal} {\bibinfo  {journal}
  {Physical review letters}\ }\textbf {\bibinfo {volume} {95}},\ \bibinfo
  {pages} {218501} (\bibinfo {year} {2005})}\BibitemShut {NoStop}%
\bibitem [{\citenamefont {Durin}\ and\ \citenamefont
  {Zapperi}(2000)}]{zapperi_prl_2000}%
  \BibitemOpen
  \bibfield  {author} {\bibinfo {author} {\bibfnamefont {G.}~\bibnamefont
  {Durin}}\ and\ \bibinfo {author} {\bibfnamefont {S.}~\bibnamefont
  {Zapperi}},\ }\href {\doibase 10.1103/PhysRevLett.84.4705} {\bibfield
  {journal} {\bibinfo  {journal} {Phys. Rev. Lett.}\ }\textbf {\bibinfo
  {volume} {84}},\ \bibinfo {pages} {4705} (\bibinfo {year}
  {2000})}\BibitemShut {NoStop}%
\bibitem [{\citenamefont {Papanikolaou}\ \emph {et~al.}(2011)\citenamefont
  {Papanikolaou}, \citenamefont {Bohn}, \citenamefont {Sommer}, \citenamefont
  {Durin}, \citenamefont {Zapperi},\ and\ \citenamefont
  {Sethna}}]{sethna_nphys_2011}%
  \BibitemOpen
  \bibfield  {author} {\bibinfo {author} {\bibfnamefont {S.}~\bibnamefont
  {Papanikolaou}}, \bibinfo {author} {\bibfnamefont {F.}~\bibnamefont {Bohn}},
  \bibinfo {author} {\bibfnamefont {R.~L.}\ \bibnamefont {Sommer}}, \bibinfo
  {author} {\bibfnamefont {G.}~\bibnamefont {Durin}}, \bibinfo {author}
  {\bibfnamefont {S.}~\bibnamefont {Zapperi}}, \ and\ \bibinfo {author}
  {\bibfnamefont {J.~P.}\ \bibnamefont {Sethna}},\ }\href@noop {} {\bibfield
  {journal} {\bibinfo  {journal} {Nature Physics}\ }\textbf {\bibinfo {volume}
  {7}},\ \bibinfo {pages} {316} (\bibinfo {year} {2011})}\BibitemShut {NoStop}%
\bibitem [{\citenamefont {{Dobrinevski}}(2013)}]{dobrinevski_thesis}%
  \BibitemOpen
  \bibfield  {author} {\bibinfo {author} {\bibfnamefont {A.}~\bibnamefont
  {{Dobrinevski}}},\ }\emph {\bibinfo {title} {{Field theory of disordered
  systems -- Avalanches of an elastic interface in a random medium}}},\
  \href@noop {} {Ph.D. thesis},\ \bibinfo  {school} {\'Ecole Normale
  Sup\'erieure, Paris} (\bibinfo {year} {2013}),\ \Eprint
  {http://arxiv.org/abs/1312.7156} {arXiv:1312.7156 [cond-mat.dis-nn]}
  \BibitemShut {NoStop}%
\bibitem [{\citenamefont {White}\ and\ \citenamefont
  {Dahmen}(2003)}]{white_prl_2003}%
  \BibitemOpen
  \bibfield  {author} {\bibinfo {author} {\bibfnamefont {R.~A.}\ \bibnamefont
  {White}}\ and\ \bibinfo {author} {\bibfnamefont {K.~A.}\ \bibnamefont
  {Dahmen}},\ }\href@noop {} {\bibfield  {journal} {\bibinfo  {journal}
  {Physical review letters}\ }\textbf {\bibinfo {volume} {91}},\ \bibinfo
  {pages} {085702} (\bibinfo {year} {2003})}\BibitemShut {NoStop}%
\bibitem [{\citenamefont {Lin}\ \emph {et~al.}(2014{\natexlab{a}})\citenamefont
  {Lin}, \citenamefont {Saade}, \citenamefont {Lerner}, \citenamefont {Rosso},\
  and\ \citenamefont {Wyart}}]{wyart_epl_2014}%
  \BibitemOpen
  \bibfield  {author} {\bibinfo {author} {\bibfnamefont {J.}~\bibnamefont
  {Lin}}, \bibinfo {author} {\bibfnamefont {A.}~\bibnamefont {Saade}}, \bibinfo
  {author} {\bibfnamefont {E.}~\bibnamefont {Lerner}}, \bibinfo {author}
  {\bibfnamefont {A.}~\bibnamefont {Rosso}}, \ and\ \bibinfo {author}
  {\bibfnamefont {M.}~\bibnamefont {Wyart}},\ }\href@noop {} {\bibfield
  {journal} {\bibinfo  {journal} {EPL (Europhysics Letters)}\ }\textbf
  {\bibinfo {volume} {105}},\ \bibinfo {pages} {26003} (\bibinfo {year}
  {2014}{\natexlab{a}})}\BibitemShut {NoStop}%
\bibitem [{\citenamefont {Karmakar}\ \emph {et~al.}(2010)\citenamefont
  {Karmakar}, \citenamefont {Lerner},\ and\ \citenamefont
  {Procaccia}}]{karmakar_pre_2010}%
  \BibitemOpen
  \bibfield  {author} {\bibinfo {author} {\bibfnamefont {S.}~\bibnamefont
  {Karmakar}}, \bibinfo {author} {\bibfnamefont {E.}~\bibnamefont {Lerner}}, \
  and\ \bibinfo {author} {\bibfnamefont {I.}~\bibnamefont {Procaccia}},\ }\href
  {\doibase 10.1103/PhysRevE.82.055103} {\bibfield  {journal} {\bibinfo
  {journal} {Phys. Rev. E}\ }\textbf {\bibinfo {volume} {82}},\ \bibinfo
  {pages} {055103} (\bibinfo {year} {2010})}\BibitemShut {NoStop}%
\bibitem [{\citenamefont {M{\"u}ller}\ and\ \citenamefont
  {Wyart}(2015)}]{wyart_anrev_2015}%
  \BibitemOpen
  \bibfield  {author} {\bibinfo {author} {\bibfnamefont {M.}~\bibnamefont
  {M{\"u}ller}}\ and\ \bibinfo {author} {\bibfnamefont {M.}~\bibnamefont
  {Wyart}},\ }\href@noop {} {\bibfield  {journal} {\bibinfo  {journal} {Annu.
  Rev. Condens. Matter Phys.}\ }\textbf {\bibinfo {volume} {6}},\ \bibinfo
  {pages} {177} (\bibinfo {year} {2015})}\BibitemShut {NoStop}%
\bibitem [{\citenamefont {Lin}\ \emph {et~al.}(2014{\natexlab{b}})\citenamefont
  {Lin}, \citenamefont {Lerner}, \citenamefont {Rosso},\ and\ \citenamefont
  {Wyart}}]{wyart_pnas_2014}%
  \BibitemOpen
  \bibfield  {author} {\bibinfo {author} {\bibfnamefont {J.}~\bibnamefont
  {Lin}}, \bibinfo {author} {\bibfnamefont {E.}~\bibnamefont {Lerner}},
  \bibinfo {author} {\bibfnamefont {A.}~\bibnamefont {Rosso}}, \ and\ \bibinfo
  {author} {\bibfnamefont {M.}~\bibnamefont {Wyart}},\ }\href@noop {}
  {\bibfield  {journal} {\bibinfo  {journal} {Proceedings of the National
  Academy of Sciences}\ }\textbf {\bibinfo {volume} {111}},\ \bibinfo {pages}
  {14382} (\bibinfo {year} {2014}{\natexlab{b}})}\BibitemShut {NoStop}%
\bibitem [{\citenamefont {Wyart}(2012)}]{wyart_prl_2012}%
  \BibitemOpen
  \bibfield  {author} {\bibinfo {author} {\bibfnamefont {M.}~\bibnamefont
  {Wyart}},\ }\href@noop {} {\bibfield  {journal} {\bibinfo  {journal}
  {Physical review letters}\ }\textbf {\bibinfo {volume} {109}},\ \bibinfo
  {pages} {125502} (\bibinfo {year} {2012})}\BibitemShut {NoStop}%
\bibitem [{\citenamefont {Imry}\ and\ \citenamefont
  {Ma}(1975)}]{imry_ma_prl_l975}%
  \BibitemOpen
  \bibfield  {author} {\bibinfo {author} {\bibfnamefont {Y.}~\bibnamefont
  {Imry}}\ and\ \bibinfo {author} {\bibfnamefont {S.-k.}\ \bibnamefont {Ma}},\
  }\href {\doibase 10.1103/PhysRevLett.35.1399} {\bibfield  {journal} {\bibinfo
   {journal} {Phys. Rev. Lett.}\ }\textbf {\bibinfo {volume} {35}},\ \bibinfo
  {pages} {1399} (\bibinfo {year} {1975})}\BibitemShut {NoStop}%
\bibitem [{\citenamefont {Hentschel}\ \emph {et~al.}(2015)\citenamefont
  {Hentschel}, \citenamefont {Jaiswal}, \citenamefont {Procaccia},\ and\
  \citenamefont {Sastry}}]{itamar_sastry_pre_2015}%
  \BibitemOpen
  \bibfield  {author} {\bibinfo {author} {\bibfnamefont {H.~G.~E.}\
  \bibnamefont {Hentschel}}, \bibinfo {author} {\bibfnamefont {P.~K.}\
  \bibnamefont {Jaiswal}}, \bibinfo {author} {\bibfnamefont {I.}~\bibnamefont
  {Procaccia}}, \ and\ \bibinfo {author} {\bibfnamefont {S.}~\bibnamefont
  {Sastry}},\ }\href {\doibase 10.1103/PhysRevE.92.062302} {\bibfield
  {journal} {\bibinfo  {journal} {Phys. Rev. E}\ }\textbf {\bibinfo {volume}
  {92}},\ \bibinfo {pages} {062302} (\bibinfo {year} {2015})}\BibitemShut
  {NoStop}%
\bibitem [{\citenamefont {{Daley}}\ and\ \citenamefont
  {{Vere-Jones}}(2008)}]{book_daley_verejones}%
  \BibitemOpen
  \bibfield  {author} {\bibinfo {author} {\bibfnamefont {D.~J.}\ \bibnamefont
  {{Daley}}}\ and\ \bibinfo {author} {\bibfnamefont {D.}~\bibnamefont
  {{Vere-Jones}}},\ }\href {\doibase 10.1007/978-0-387-49835-5} {\emph
  {\bibinfo {title} {An introduction to the theory of point processes. Vol. II:
  General theory and structure.}}},\ \bibinfo {edition} {2nd}\ ed.\ (\bibinfo
  {publisher} {New York, NY: Springer},\ \bibinfo {year} {2008})\ pp.\ \bibinfo
  {pages} {xvii + 573}\BibitemShut {NoStop}%
\bibitem [{\citenamefont {{Sabhapandit}}(2002)}]{sanjib_thesis}%
  \BibitemOpen
  \bibfield  {author} {\bibinfo {author} {\bibfnamefont {S.}~\bibnamefont
  {{Sabhapandit}}},\ }\emph {\bibinfo {title} {{Hysteresis and Avalanches in
  the Random Field Ising Model}}},\ \href@noop {} {Ph.D. thesis},\ \bibinfo
  {school} {Tata Institute of Fundamental Research, Mumbai} (\bibinfo {year}
  {2002})\BibitemShut {NoStop}%
\bibitem [{\citenamefont {Sethna}\ \emph {et~al.}(1993)\citenamefont {Sethna},
  \citenamefont {Dahmen}, \citenamefont {Kartha}, \citenamefont {Krumhansl},
  \citenamefont {Roberts},\ and\ \citenamefont {Shore}}]{sethna_prl_1993}%
  \BibitemOpen
  \bibfield  {author} {\bibinfo {author} {\bibfnamefont {J.~P.}\ \bibnamefont
  {Sethna}}, \bibinfo {author} {\bibfnamefont {K.}~\bibnamefont {Dahmen}},
  \bibinfo {author} {\bibfnamefont {S.}~\bibnamefont {Kartha}}, \bibinfo
  {author} {\bibfnamefont {J.~A.}\ \bibnamefont {Krumhansl}}, \bibinfo {author}
  {\bibfnamefont {B.~W.}\ \bibnamefont {Roberts}}, \ and\ \bibinfo {author}
  {\bibfnamefont {J.~D.}\ \bibnamefont {Shore}},\ }\href {\doibase
  10.1103/PhysRevLett.70.3347} {\bibfield  {journal} {\bibinfo  {journal}
  {Phys. Rev. Lett.}\ }\textbf {\bibinfo {volume} {70}},\ \bibinfo {pages}
  {3347} (\bibinfo {year} {1993})}\BibitemShut {NoStop}%
\bibitem [{\citenamefont {Grinstein}\ and\ \citenamefont
  {Ma}(1983)}]{grinstein_prb_1983}%
  \BibitemOpen
  \bibfield  {author} {\bibinfo {author} {\bibfnamefont {G.}~\bibnamefont
  {Grinstein}}\ and\ \bibinfo {author} {\bibfnamefont {S.-k.}\ \bibnamefont
  {Ma}},\ }\href@noop {} {\bibfield  {journal} {\bibinfo  {journal} {Physical
  Review B}\ }\textbf {\bibinfo {volume} {28}},\ \bibinfo {pages} {2588}
  (\bibinfo {year} {1983})}\BibitemShut {NoStop}%
\bibitem [{\citenamefont {Middleton}(1992)}]{middleton_prl_1992}%
  \BibitemOpen
  \bibfield  {author} {\bibinfo {author} {\bibfnamefont {A.~A.}\ \bibnamefont
  {Middleton}},\ }\href {\doibase 10.1103/PhysRevLett.68.670} {\bibfield
  {journal} {\bibinfo  {journal} {Phys. Rev. Lett.}\ }\textbf {\bibinfo
  {volume} {68}},\ \bibinfo {pages} {670} (\bibinfo {year} {1992})}\BibitemShut
  {NoStop}%
\bibitem [{\citenamefont {Dhar}\ \emph {et~al.}(1997)\citenamefont {Dhar},
  \citenamefont {Shukla},\ and\ \citenamefont {Sethna}}]{dhar_jphysa_1997}%
  \BibitemOpen
  \bibfield  {author} {\bibinfo {author} {\bibfnamefont {D.}~\bibnamefont
  {Dhar}}, \bibinfo {author} {\bibfnamefont {P.}~\bibnamefont {Shukla}}, \ and\
  \bibinfo {author} {\bibfnamefont {J.~P.}\ \bibnamefont {Sethna}},\ }\href
  {http://stacks.iop.org/0305-4470/30/i=15/a=013} {\bibfield  {journal}
  {\bibinfo  {journal} {Journal of Physics A: Mathematical and General}\
  }\textbf {\bibinfo {volume} {30}},\ \bibinfo {pages} {5259} (\bibinfo {year}
  {1997})}\BibitemShut {NoStop}%
\bibitem [{\citenamefont {Sabhapandit}\ \emph {et~al.}(2000)\citenamefont
  {Sabhapandit}, \citenamefont {Shukla},\ and\ \citenamefont
  {Dhar}}]{sanjib_jstatphys_2000}%
  \BibitemOpen
  \bibfield  {author} {\bibinfo {author} {\bibfnamefont {S.}~\bibnamefont
  {Sabhapandit}}, \bibinfo {author} {\bibfnamefont {P.}~\bibnamefont {Shukla}},
  \ and\ \bibinfo {author} {\bibfnamefont {D.}~\bibnamefont {Dhar}},\
  }\href@noop {} {\bibfield  {journal} {\bibinfo  {journal} {Journal of
  Statistical Physics}\ }\textbf {\bibinfo {volume} {98}},\ \bibinfo {pages}
  {103} (\bibinfo {year} {2000})}\BibitemShut {NoStop}%
\bibitem [{\citenamefont {Bruinsma}\ and\ \citenamefont
  {Aeppli}(1984)}]{bruinsma_prl_1984}%
  \BibitemOpen
  \bibfield  {author} {\bibinfo {author} {\bibfnamefont {R.}~\bibnamefont
  {Bruinsma}}\ and\ \bibinfo {author} {\bibfnamefont {G.}~\bibnamefont
  {Aeppli}},\ }\href@noop {} {\bibfield  {journal} {\bibinfo  {journal}
  {Physical review letters}\ }\textbf {\bibinfo {volume} {52}},\ \bibinfo
  {pages} {1547} (\bibinfo {year} {1984})}\BibitemShut {NoStop}%
\bibitem [{\citenamefont {Travesset}\ \emph {et~al.}(2002)\citenamefont
  {Travesset}, \citenamefont {White},\ and\ \citenamefont
  {Dahmen}}]{travesset_prb_2002}%
  \BibitemOpen
  \bibfield  {author} {\bibinfo {author} {\bibfnamefont {A.}~\bibnamefont
  {Travesset}}, \bibinfo {author} {\bibfnamefont {R.~A.}\ \bibnamefont
  {White}}, \ and\ \bibinfo {author} {\bibfnamefont {K.~A.}\ \bibnamefont
  {Dahmen}},\ }\href@noop {} {\bibfield  {journal} {\bibinfo  {journal}
  {Physical Review B}\ }\textbf {\bibinfo {volume} {66}},\ \bibinfo {pages}
  {024430} (\bibinfo {year} {2002})}\BibitemShut {NoStop}%
\bibitem [{\citenamefont {Andresen}\ \emph {et~al.}(2016)\citenamefont
  {Andresen}, \citenamefont {Pramudya}, \citenamefont {Katzgraber},
  \citenamefont {Thomas}, \citenamefont {Zimanyi},\ and\ \citenamefont
  {Dobrosavljevi{\'c}}}]{andresen2016charge}%
  \BibitemOpen
  \bibfield  {author} {\bibinfo {author} {\bibfnamefont {J.~C.}\ \bibnamefont
  {Andresen}}, \bibinfo {author} {\bibfnamefont {Y.}~\bibnamefont {Pramudya}},
  \bibinfo {author} {\bibfnamefont {H.~G.}\ \bibnamefont {Katzgraber}},
  \bibinfo {author} {\bibfnamefont {C.~K.}\ \bibnamefont {Thomas}}, \bibinfo
  {author} {\bibfnamefont {G.~T.}\ \bibnamefont {Zimanyi}}, \ and\ \bibinfo
  {author} {\bibfnamefont {V.}~\bibnamefont {Dobrosavljevi{\'c}}},\ }\href@noop
  {} {\bibfield  {journal} {\bibinfo  {journal} {Physical Review B}\ }\textbf
  {\bibinfo {volume} {93}},\ \bibinfo {pages} {094429} (\bibinfo {year}
  {2016})}\BibitemShut {NoStop}%
\bibitem [{map()}]{mapping}%
  \BibitemOpen
  \href@noop {} {}\bibinfo {note} {The mapping of the Coulomb glass Hamiltonian
  $\mathcal{H} = \sum_{i} \sum_{j} \frac{(n_i - \langle n \rangle) (n_j -
  \langle n \rangle)}{|i - j|} + \sum_{i} \epsilon_i n_i$, with $n_i = 0,1$ and
  $\epsilon$'s representing random on-site energies to a model of Ising spins
  with $S_i = 2 n_i - 1$ introduces an $M^2$ term, where $M = \frac{1}{N}
  \sum_{i}^{N} S_i$.}\BibitemShut {Stop}%
\bibitem [{\citenamefont {Kerimov}(1993)}]{kerimov_jstatphys_1993}%
  \BibitemOpen
  \bibfield  {author} {\bibinfo {author} {\bibfnamefont {A.}~\bibnamefont
  {Kerimov}},\ }\href@noop {} {\bibfield  {journal} {\bibinfo  {journal}
  {Journal of statistical physics}\ }\textbf {\bibinfo {volume} {72}},\
  \bibinfo {pages} {571} (\bibinfo {year} {1993})}\BibitemShut {NoStop}%
\bibitem [{\citenamefont {Tadi\ifmmode~\acute{c}\else \'{c}\fi{}}\ \emph
  {et~al.}(2005)\citenamefont {Tadi\ifmmode~\acute{c}\else \'{c}\fi{}},
  \citenamefont {Malarz},\ and\ \citenamefont
  {Ku\l{}akowski}}]{tadic_prl_2005}%
  \BibitemOpen
  \bibfield  {author} {\bibinfo {author} {\bibfnamefont {B.}~\bibnamefont
  {Tadi\ifmmode~\acute{c}\else \'{c}\fi{}}}, \bibinfo {author} {\bibfnamefont
  {K.}~\bibnamefont {Malarz}}, \ and\ \bibinfo {author} {\bibfnamefont
  {K.}~\bibnamefont {Ku\l{}akowski}},\ }\href {\doibase
  10.1103/PhysRevLett.94.137204} {\bibfield  {journal} {\bibinfo  {journal}
  {Phys. Rev. Lett.}\ }\textbf {\bibinfo {volume} {94}},\ \bibinfo {pages}
  {137204} (\bibinfo {year} {2005})}\BibitemShut {NoStop}%
\bibitem [{\citenamefont {Kurbah}\ and\ \citenamefont
  {Shukla}(2011)}]{shukla_pre_2011}%
  \BibitemOpen
  \bibfield  {author} {\bibinfo {author} {\bibfnamefont {L.}~\bibnamefont
  {Kurbah}}\ and\ \bibinfo {author} {\bibfnamefont {P.}~\bibnamefont
  {Shukla}},\ }\href {\doibase 10.1103/PhysRevE.83.061136} {\bibfield
  {journal} {\bibinfo  {journal} {Phys. Rev. E}\ }\textbf {\bibinfo {volume}
  {83}},\ \bibinfo {pages} {061136} (\bibinfo {year} {2011})}\BibitemShut
  {NoStop}%
\bibitem [{\citenamefont {Palassini}\ and\ \citenamefont
  {Goethe}(2012)}]{palassini2012elementary}%
  \BibitemOpen
  \bibfield  {author} {\bibinfo {author} {\bibfnamefont {M.}~\bibnamefont
  {Palassini}}\ and\ \bibinfo {author} {\bibfnamefont {M.}~\bibnamefont
  {Goethe}},\ }in\ \href@noop {} {\emph {\bibinfo {booktitle} {Journal of
  Physics: Conference Series}}},\ Vol.\ \bibinfo {volume} {376}\ (\bibinfo
  {organization} {IOP Publishing},\ \bibinfo {year} {2012})\ p.\ \bibinfo
  {pages} {012009}\BibitemShut {NoStop}%
\bibitem [{\citenamefont {Tadi\'c}(2000)}]{tadic_physica_A_2000}%
  \BibitemOpen
  \bibfield  {author} {\bibinfo {author} {\bibfnamefont {B.}~\bibnamefont
  {Tadi\'c}},\ }\href {\doibase
  http://dx.doi.org/10.1016/S0378-4371(00)00099-6} {\bibfield  {journal}
  {\bibinfo  {journal} {Physica A: Statistical Mechanics and its Applications}\
  }\textbf {\bibinfo {volume} {282}},\ \bibinfo {pages} {362 } (\bibinfo {year}
  {2000})}\BibitemShut {NoStop}%
\bibitem [{\citenamefont {Yakovlev}\ \emph {et~al.}(2005)\citenamefont
  {Yakovlev}, \citenamefont {Rundle}, \citenamefont {Shcherbakov},\ and\
  \citenamefont {Turcotte}}]{yakovlev_arxiv_2005}%
  \BibitemOpen
  \bibfield  {author} {\bibinfo {author} {\bibfnamefont {G.}~\bibnamefont
  {Yakovlev}}, \bibinfo {author} {\bibfnamefont {J.~B.}\ \bibnamefont
  {Rundle}}, \bibinfo {author} {\bibfnamefont {R.}~\bibnamefont {Shcherbakov}},
  \ and\ \bibinfo {author} {\bibfnamefont {D.~L.}\ \bibnamefont {Turcotte}},\
  }\href@noop {} {\bibfield  {journal} {\bibinfo  {journal} {arXiv preprint
  cond-mat/0507657}\ } (\bibinfo {year} {2005})}\BibitemShut {NoStop}%
\end{thebibliography}%
\bibliographystyle{apsrev4-1}

\end{document}